\documentclass[10pt,conference]{IEEEtran}
\usepackage{cite}
\usepackage{amsmath,amssymb,amsfonts}
\usepackage{algorithmic}
\usepackage{graphicx}
\usepackage{textcomp}
\usepackage{xcolor}
\usepackage[hyphens]{url}
\usepackage{fancyhdr}
\usepackage{hyperref}

\usepackage{lipsum}

\usepackage{multirow}
\usepackage{multicol}
\usepackage{booktabs}
\usepackage{xspace}
\usepackage[T1]{fontenc}

\newcommand{\Fig}[1]{Fig.~\ref{#1}}

\newcommand{\Tbl}[1]{Tbl.~\ref{#1}}
\newcommand{\Sec}[1]{Sec.~\ref{#1}}
\newcommand{\Reb}[1]{\textcolor{red}}

\newcommand{\proj}{\textit{Focus}\xspace}

\newcommand{\hpcayear}{2026}

\newcommand{\hpcasubmissionnumber}{4}
\title{Focus: A Streaming Concentration Architecture for Efficient Vision-Language Models}

\def\hpcacameraready{} 

\newcommand\hpcaauthors{
Chiyue Wei$^{\dagger*}$,
Cong Guo$^{\dagger*\S}$,
Junyao Zhang$^{\dagger}$,
Haoxuan Shan$^{\dagger}$,
Yifan Xu$^{\dagger}$,
Ziyue Zhang$^{\dagger}$, \\
Yudong Liu$^{\dagger}$,
Qinsi Wang$^{\dagger}$,
Changchun Zhou$^{\dagger}$,
Hai ``Helen'' Li$^{\dagger}$,
Yiran Chen$^{\dagger}$
}

\newcommand\hpcaaffiliation{
$^{\dagger}$Duke University \quad
$^{*}$Equal contribution \quad
$^{\S}$Corresponding author
}

\newcommand\hpcaemail{
\{chiyue.wei, cong.guo, hai.li, yiran.chen\}@duke.edu
}

\author{
  \ifdefined\hpcacameraready
    \IEEEauthorblockN{\hpcaauthors{}}
      \IEEEauthorblockA{
        \hpcaaffiliation{} \\
        \hpcaemail{}
      }
  \else
    \IEEEauthorblockN{\normalsize{HPCA \hpcayear{} Submission
      \textbf{\#\hpcasubmissionnumber{}}} \\
      \IEEEauthorblockA{
        Confidential Draft \\
        Do NOT Distribute!!
      }
    }
  \fi 
}

\fancypagestyle{camerareadyfirstpage}{%
  \fancyhead{}
  
  \fancyhead[C]{
    \ifdefined\aeopen
    \parbox[][12mm][t]{13.5cm}{\hpcayear{} IEEE International Symposium on High-Performance Computer Architecture (HPCA)}    
    \else
      \ifdefined\aereviewed
      \parbox[][12mm][t]{13.5cm}{\hpcayear{} IEEE International Symposium on High-Performance Computer Architecture (HPCA)}
      \else
      \ifdefined\aereproduced
      \parbox[][12mm][t]{13.5cm}{\hpcayear{} IEEE International Symposium on High-Performance Computer Architecture (HPCA)}
      \else
      \parbox[][0mm][t]{13.5cm}{\hpcayear{} IEEE International Symposium on High-Performance Computer Architecture (HPCA)}
    \fi 
    \fi 
    \fi 
    \ifdefined\aeopen 
      \includegraphics[width=12mm,height=12mm]{ae-badges/open-research-objects.pdf}
    \fi 
    \ifdefined\aereviewed
      \includegraphics[width=12mm,height=12mm]{ae-badges/research-objects-reviewed.pdf}
    \fi 
    \ifdefined\aereproduced
      \includegraphics[width=12mm,height=12mm]{ae-badges/results-reproduced.pdf}
    \fi
  }
  \fancyfoot[C]{}
}
\begin{document}
\maketitle

\ifdefined\hpcacameraready 
  \thispagestyle{camerareadyfirstpage}
  \pagestyle{empty}
\else
  \thispagestyle{plain}
  \pagestyle{plain}
\fi

\newcommand{\hpcaheight}{0mm}
\ifdefined\eaopen
\renewcommand{\hpcaheight}{12mm}
\fi

\begin{abstract}
Vision-Language Models (VLMs) have demonstrated strong performance on tasks such as video captioning and visual question answering. 
However, their growing scale and video-level inputs lead to significant computational and memory overhead, posing challenges for real-time deployment on hardware accelerators. 
While prior work attempts to reduce redundancy via token pruning or merging, these methods typically operate at coarse granularity and incur high runtime overhead due to global token-level operations.

In this study, we propose \textbf{\textit{Focus}}, a \textit{Streaming Concentration Architecture} that efficiently accelerates VLM inference through progressive, fine-grained redundancy elimination. 
\textit{Focus} introduces a multilevel concentration paradigm that hierarchically compresses vision-language inputs at three levels: 
(1) semantic-guided token pruning based on textual prompts, 
(2) spatial-temporal block-level concentration using localized comparisons, 
and (3) vector-level redundancy removal via motion-aware matching.
All concentration steps are tightly co-designed with the architecture to support streaming-friendly, on-chip execution. 
\textit{Focus} leverages GEMM tiling, convolution-style layout, and cross-modal attention to minimize off-chip access while enabling high throughput. 
Implemented as a modular unit within a systolic-array accelerator, \textit{Focus} achieves {\textbf{2.4$\times$} speedup and \textbf{3.3$\times$} reduction in energy}, significantly outperforming {state-of-the-art accelerator} in both performance and energy efficiency. Full-stack implementation of \proj is open-sourced at \href{https://github.com/dubcyfor3/Focus}{https://github.com/dubcyfor3/Focus}.
\end{abstract}

\section{Introduction}
\label{sec:intro}

Vision-Language Models (VLMs)~\cite{li2024llava,liu2023llava} have emerged as a cornerstone of multimodal AI, enabling joint reasoning over visual and textual data. 
By integrating advances from computer vision and natural language processing, VLMs excel at tasks such as video captioning~\cite{yang2023vid2seq,zhang2024video}, visual question answering~\cite{sinha2024guiding_vlm_sel_for_vqa,das2025vlm_continual_vqa}, and cross-modal retrieval~\cite{li2022blip}. 
Following a similar trajectory to Large Language Models (LLMs)~\cite{brown2020language, devlin2018bert}, modern VLMs have rapidly scaled in size and data, resulting in notable accuracy gains. 
However, this scaling significantly increases compute and memory demands, posing challenges for deployment, especially on edge devices~\cite{sharshar2025vision}.

\begin{figure}
    \centering
    \includegraphics[width=1\linewidth]{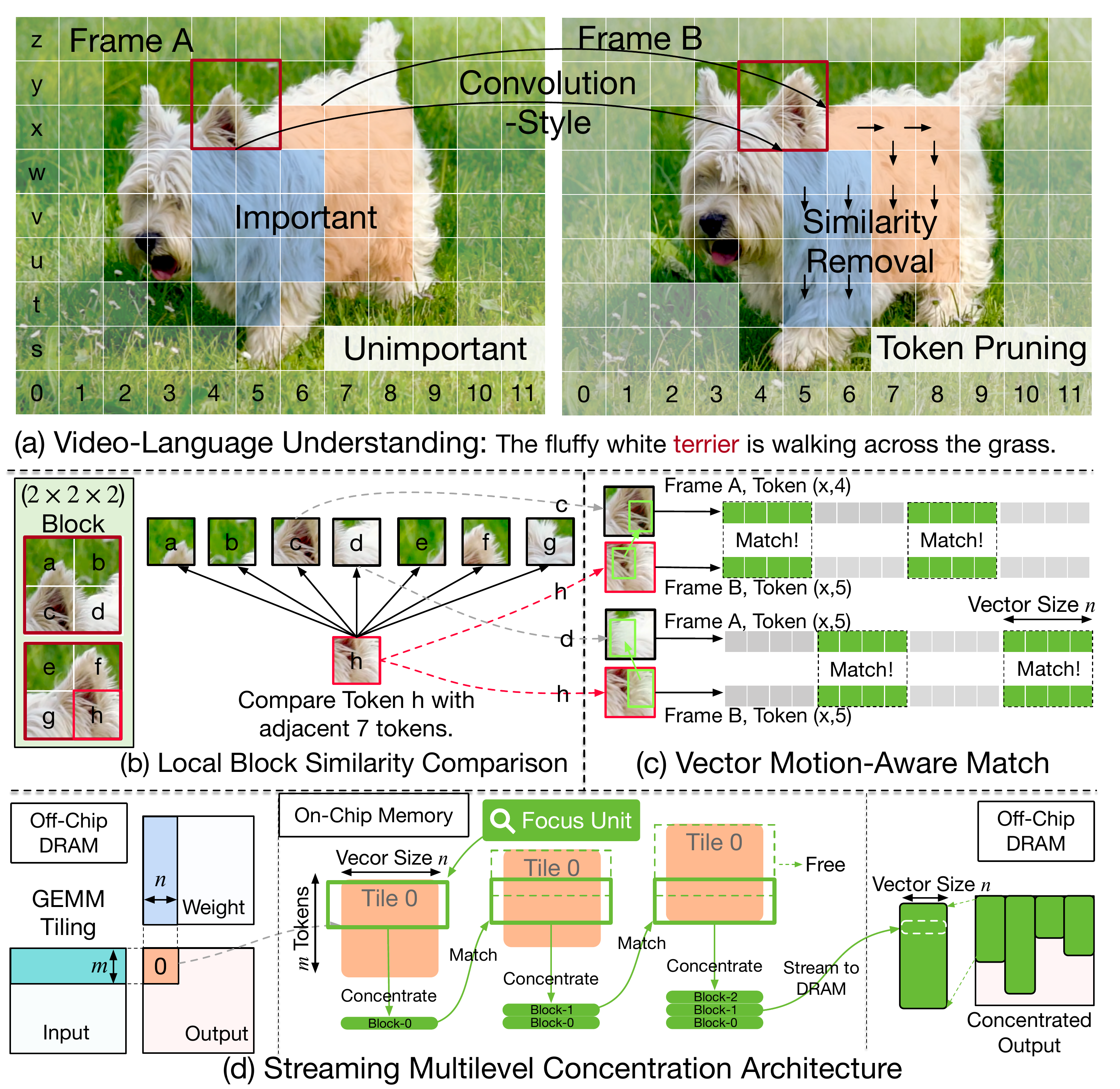}
    \caption{Overview of the streaming multilevel concentration architecture.}
    \label{fig:intro}
\end{figure}

Fortunately, video-based inputs offer a key opportunity: high visual redundancy~\cite{dynamicllava,voco,prumerge,fastv,wang2025corematching}.  
As shown in \Fig{fig:intro}(a), adjacent frames often share similar backgrounds and foreground objects.  
Since VLMs tokenize each frame independently~\cite{li2024llava,zhang2024video}, many tokens across or within frames are redundant.  
This has motivated techniques such as token pruning~\cite{sah2024token_pruning_vit,wang2025corematching} and token merging~\cite{bolya2023token} to reduce computation.  
However, most prior work focuses on algorithmic strategies without considering hardware alignment.  
For instance, Token Merging~\cite{bolya2023token} introduces a ToMe module that increases runtime by up to 36.8\%~\cite{yoo2024adaptiv}.

Recent designs such as AdapTiV~\cite{yoo2024adaptiv} and CMC~\cite{song2024cmc} address these inefficiencies at the hardware level.  
AdapTiV implements a simplified ToMe module in hardware, while CMC leverages video-codec-inspired compression (e.g., H.264~\cite{wiegand2003overview}) via an external codec block.
However, both approaches largely translate existing algorithms without embracing full hardware-{algorithm} co-design. 
First, both {targeted for Vision Transformers (ViTs)~\cite{dosovitskiy2020image}, }focus only on visual redundancy and overlook the cross-modal nature of VLMs. 
CMC’s codec ignores language inputs, and AdapTiV only supports static images, missing video-language interactions.
Second, both operate at global token-level granularity, which is inefficient for both algorithm and hardware due to high overhead and poor locality.
To enable efficient VLM deployment, a more holistic co-design approach is needed, one that leverages cross-modal redundancy while aligning with hardware-friendly processing granularity.

In this study, we propose a novel architecture, \textbf{\proj{}}, to accelerate VLM inference by performing \textbf{\textit{streaming concentration}}, a multilevel compression technique that removes visual and cross-modal redundancy in a streaming-friendly, on-chip processing fashion.

From the \textbf{algorithmic perspective}, \proj{} performs redundancy concentration at three levels of granularity.  
First, it leverages semantic understanding to retain only visual regions relevant to the textual prompt.  
Prior work~\cite{rao2021dynamicvit,bolya2023token,song2024cmc,yoo2024adaptiv} relies on static metrics like token magnitude, which fail to capture prompt-conditioned semantics in VLMs.  
As shown in \Fig{fig:intro}(a), attention may shift from a foreground object (e.g., a dog) to a background element (e.g., a flower), depending on the question (see details in \Sec{subsec:motivation_algo}).
To address this, \proj{} introduces a prompt-aware importance analyzer that dynamically prunes visual tokens based on cross-modal attention, improving both accuracy and efficiency.

Second, as illustrated in \Fig{fig:intro}(b), \proj{} groups retained tokens into spatiotemporal blocks, using the last token (e.g., token~$h$) as the key for localized similarity comparisons. 
The key token is compared with others in its block. 
This is applied across the video, treating each token in turn as a key. 
This technique resembles a 3D convolutional sweep that progressively concentrates similarity through localized matching. 
By operating within small spatial-temporal windows, \proj{} avoids global comparisons, making the process compute-efficient and highly streamable.

Third, \proj{} explores redundancy at the vector level. 
Due to video motion, a token may align with multiple shifted tokens in adjacent frames. 
As shown in \Fig{fig:intro}(c), token $h$ may share features with parts of tokens $c$ and $d$. 
Relying on a single best-matched token may lose information. 
Instead, \proj{} performs vector-wise comparisons, allowing each vector to match multiple candidates and capture richer sub-token similarity. 
By integrating these three levels, \proj{} achieves up to \textbf{83\%} (\textbf{80\%} on average) computational sparsity through multilevel concentration, significantly outperforming CMC and AdapTiV, which typically reach only 40--50\%, under similar accuracy.

From the \textbf{architectural perspective}, \proj{} is designed to efficiently support multilevel concentration through tight alignment with General Matrix Multiplication (GEMM) tiling.  
As shown in \Fig{fig:intro}(d), its vector- and block-level operations naturally align with the tiling strategies widely adopted in systolic-array–based accelerators such as TPUs~\cite{jouppi2023tpuv4} and GPUs~\cite{choquette2021nvidia}.
GEMM tiling addresses on-chip memory constraints by dividing matrices into small, independently processed tiles. 
Each tile is handled by the PE array in isolation, enabling efficient, in-place \textbf{vector-level} similarity detection and compression.
By eliminating redundancy locally within each tile, \proj{} minimizes data movement and reduces both compute cost and DRAM traffic. 
In contrast to global token-wise methods that rely on costly off-chip access, \proj{} achieves fine-grained, on-chip processing in a hardware-efficient manner.

At the \textbf{block level}, \proj{} draws inspiration from CNN accelerators~\cite{chen2016eyeriss,du2015shidiannao}, using a sliding window to stream and process output tokens directly from the compute core (e.g., systolic array), maximizing locality and sustaining high throughput. 
To handle the non-contiguous nature of VLM tokens within a block, we adopt a convolution-style layout that preserves streaming flow while ensuring alignment for block-wise matching.
At the {semantic-level, which corresponds to the \textbf{token level}}, \proj{} integrates into the attention layer to identify and retain the most relevant tokens based on cross-modal attention scores.  
Through dedicated scheduling, it performs token selection in a streaming fashion, enabling compression prior to memory write-back without stalling GEMM execution.

\proj{} operates as a standalone module, similar to pooling or activation, without interfering with the core computation pipeline.  
Its modularity enables broad applicability and scalability while maintaining high compression efficiency.  
By co-optimizing the algorithm and architecture, \proj{} achieves up to \textbf{5.0$\times$} reduction in computation and \textbf{4.5$\times$} reduction in memory footprint for VLM inference.
Occupying only {\textbf{2.7\%}} of the systolic array area, it is lightweight and well-suited for edge deployment.
{Our contributions are as follows:}
\begin{itemize}
    \item We propose {multilevel concentration}, a hardware-oriented redundancy removal paradigm that eliminates semantic-, block-, and vector-level redundancy in VLMs.

    \item We develop a co-designed {streaming concentration architecture} that aligns with tiling-based execution and memory access patterns with minimal hardware overhead.

    \item To the best of our knowledge, \proj is the first architecture tailored for VLMs, delivering {2.60$\times$}/{2.35$\times$} performance and {{2.98$\times$}/{3.29$\times$}} energy efficiency gains over AdapTiV and CMC, respectively.
\end{itemize}

\section{Background}

\subsection{Vision-Language Models}
The success of Large Language Models (LLMs), such as GPT~\cite{achiam2023gpt} and LLaMA~\cite{touvron2023llama}, has driven remarkable progress across a broad range of applications. Building upon this foundation, Vision-Language Models (VLMs) extend the capabilities of LLMs to multimodal inputs, enabling joint reasoning over visual and textual information. This multimodal capability significantly broadens their utility in tasks such as video captioning~\cite{maeda2024vlm_based_caption_eval}, visual question answering (VQA)~\cite{das2025vlm_continual_vqa,sinha2024guiding_vlm_sel_for_vqa}, and interactive multimodal assistants~\cite{liu2023llava,bordes2024introduction}. 
With superior adaptability and generalization in open-world visual scenarios, VLMs are emerging as a transformative technology with far-reaching impact in both academic~\cite{alayrac2022flamingo,lin2024vila,girdhar2023imagebind} and industrial~\cite{achiam2023gpt,team2024gemini,aws2023titan_multimodal_embeddings} domains.

Modern VLMs consist of a vision encoder and a Large Language Model (LLM) that jointly process visual and textual inputs. 
In video-based VLMs, videos are sampled into frames, divided into patches, and tokenized by the vision encoder into embeddings, which are projected into the LLM’s word embedding space for multimodal fusion. 
These visual tokens are concatenated with text prompts and processed by the LLM to generate text outputs.

The LLM with Transformer~\cite{vaswani2017attention,brown2020language} model architecture dominates both model size and computation. 
For example, in LLaVA-OneVision-72B~\cite{li2024llava}, it accounts for \textbf{99.35\%} of parameters and \textbf{98.98\%} of operations. 
{Moreover, visual tokens typically make up 98\%–99\% of total inputs; in LLaVA-OneVision on the VideoMME dataset~\cite{fu2025video}, each sample averages 6,272 visual tokens versus only 109 text tokens.} 
Therefore, optimizing LLM efficiency is crucial for accelerating VLMs.

\subsection{Efficiency Optimizations for VLM}

\textbf{Efficient Algorithms.}  
Video-based VLMs generate a large number of tokens, placing heavy demands on compute and memory. To mitigate this, various token pruning techniques have been proposed~\cite{dynamicllava,voco,fastv,liu2025keyframe,wang2025corematching}. 
For instance, Prumerge~\cite{prumerge} uses sparse attention scores between the class token and visual tokens to discard less important ones, while FrameFusion~\cite{fu2024framefusion} merges temporally redundant tokens across frames.

These methods show that only a small subset of tokens is needed to preserve performance. However, they often incur runtime overhead for importance estimation and produce irregular sparsity patterns that limit GPU utilization.

\textbf{Hardware Accelerators.}  
Dedicated VLM accelerators are still rare, though Vision Transformer (ViT) accelerators~\cite{song2024cmc,yoo2024adaptiv,dong2023heatvit} offer transferable insights.  
AdapTiV~\cite{yoo2024adaptiv} merges nearby tokens using lightweight similarity checks based on sign bits, while CMC~\cite{song2024cmc} leverages video codec hardware to detect inter-frame redundancy.
These designs offload token selection to specialized logic, reducing overhead and enabling efficient sparsity utilization, but are limited to coarse-grained, token-level pruning.

In contrast, \proj{} captures both coarse- and fine-grained redundancy through a multi-level concentration strategy. This broadens the scope of efficiency gains and enables hardware-friendly sparsity, as detailed in the following sections.

\begin{figure}[t]
    \centering
    \includegraphics[width=1\linewidth]{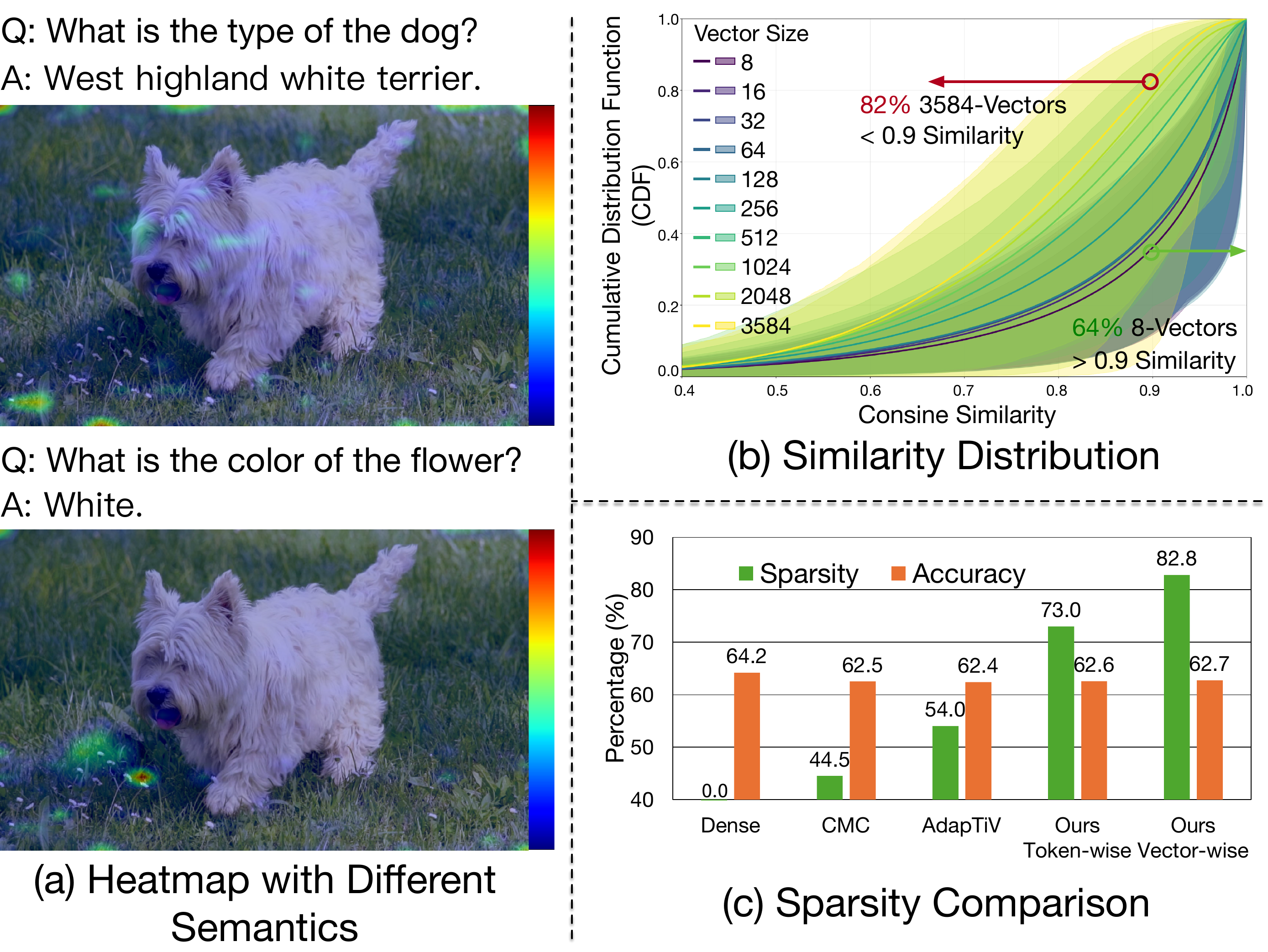}
    \caption{Motivation for multilevel concentration. 
    {(a) Prompt-aware attention heatmaps.} 
    {(b) Cosine similarity CDFs.} 
    (c) Sparsity Comparison.}
    \label{fig:motivation1}
\end{figure}

\section{Motivation}
This section presents our motivation from both algorithmic and architectural perspectives across three levels:

(1) \textit{Token (semantic) level} prunes irrelevant tokens based on language context through semantic concentration;

(2) \textit{Block level (similarity scope)}  detects local spatiotemporal redundancy within adjacent regions using block-wise comparison;

(3) \textit{Vector level (similarity granularity)} captures fine-grained sub-token redundancy via vector-wise similarity.

\subsection{Algorithm: Multilevel Concentration}
\label{subsec:motivation_algo}
\textbf{Semantic Attention Shifts with the Prompt.}
In Vision-Language Models (VLMs), token importance is inherently tied to the input prompt. 
Prior pruning methods often rely on static heuristics such as saliency or token magnitude, which fail to capture prompt-specific semantic intent.

To illustrate this, we extract cross-modal attention maps {averaged from all layers} of the Llava-Onevision-7B~\cite{li2024llava} model under two different prompts, as shown in \Fig{fig:motivation1}(a). 
When asked \textit{``What is the type of the dog?''}, attention concentrates on the dog; when asked \textit{``What is the color of the flower?''}, attention shifts to the lower-left corner where the flowers reside. 
These examples highlight that semantically relevant tokens vary greatly with the question, and static importance metrics are inadequate.

Our semantic concentration module leverages cross-modal attention to prune uninformative tokens early, improving efficiency without degrading accuracy.

\textbf{Fine-grained Granularity Enhances Redundancy Detection.}  
Global token-level matching is often too coarse to capture redundancy arising from motion, deformation, or soft spatial shifts. 
As illustrated in \Fig{fig:intro}(c), a token in one frame may partially overlap with several neighboring tokens in the next frame, rendering single-token matching ineffective.

To better capture such partial alignments, we divide token embeddings into {vectors} and perform similarity comparisons at the vector level.
We extract {all layers' input} from Llava-OneVision~\cite{li2024llava} model with the MLVU~\cite{zhou2025mlvu} dataset.
{\Fig{fig:motivation1}(b) shows the average cosine similarity distribution across all layers, along with the variation range among layers. On average, over 64\% of 8-dimensional vectors exceed a cosine similarity threshold of 0.9, compared to only 18\% for 3584-dimensional vectors, indicating that finer granularity reveals substantially more redundancy.}
This enables higher sparsity without degrading accuracy.  
As shown in \Fig{fig:motivation1}(c), our vector-level method achieves 82.8\% sparsity on Llava-Video~\cite{zhang2024video} with the VideoMME~\cite{fu2025video} dataset, outperforming both CMC and AdapTiV, and exceeding our token-wise variant by 9.8\%. 
This translates to a 1.6$\times$ reduction in computation, while slightly improving accuracy.

Block- and vector-level strategies are complementary: block granularity defines \textit{where} comparisons are applied (e.g., within spatiotemporal windows), while vector granularity determines \textit{how fine} those comparisons are conducted.  
Together, they yield structured sparsity that aligns naturally with GEMM tiling, enabling efficient and accurate compression in hardware.

\begin{figure}[t]
    \centering
    \includegraphics[width=1\linewidth]{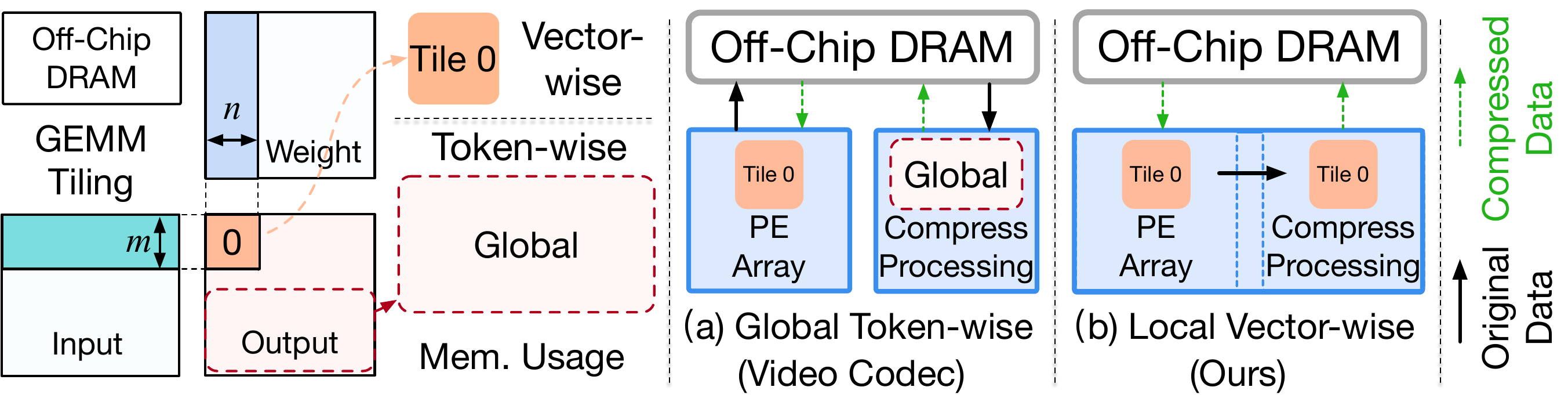}
    \caption{
    (a) Global token-wise methods (e.g., CMC) perform compression off-chip after writing all token outputs to DRAM. 
    (b) \proj{} compresses locally and on-chip at the vector level, immediately after each tile is produced. 
    }
    \label{fig:motivation2}
\end{figure}

\subsection{Architecture: Hardware-Oriented Design}
While many efforts aim to improve VLM efficiency~\cite{bolya2023token,fu2024framefusion,liu2025nvila}, most focus on algorithmic techniques while overlooking hardware constraints.  
In high-throughput systems, algorithm and architecture must be co-designed, otherwise, even efficient algorithms can suffer from memory bottlenecks or poor data locality.  
As shown in \Fig{fig:motivation2}, our design bridges this gap through a \textbf{vector-wise compression strategy} that improves both accuracy and system efficiency.

\textbf{Limitations of Global Token-Wise Methods.}  
Prior designs like CMC~\cite{song2024cmc} adopt global, token-wise compression by offloading redundancy removal to a codec unit after writing full token outputs to DRAM~(\Fig{fig:motivation2}a).  
This incurs high bandwidth usage and sacrifices data locality.  
AdapTiV~\cite{yoo2024adaptiv} integrates token merging into hardware, but still relies on coarse token-pair operations and must transfer uncompressed tokens before processing.
{If prior designs were required to perform compression before writing back to DRAM, they would need an additional large buffer; for example, CMC uses up to 1.4MB.}
Token-wise methods also require full-token readiness before redundancy detection, limiting streaming.

Moreover, these approaches overlook sub-token redundancy and operate at the full GEMM level, which misaligns with the execution model of systolic arrays that process small, regular GEMM tiles.
Their global execution prevents fine-grained scheduling and increases memory pressure.  
As shown in \Sec{sec:mem_access}, CMC achieves 46\% sparsity but still incurs 79\% of dense DRAM traffic, whereas \proj{} reaches 81\% sparsity with only 21\% of the bandwidth, highlighting the advantage of hardware-aligned, vector-level concentration.

\textbf{GEMM-Tile Friendly Compression.}  
\proj{} performs compression entirely within each GEMM tile, aligning with the compute flow of systolic arrays.  
As shown in \Fig{fig:motivation2}(b), vector-level similarity is computed immediately after generating each $m \times n$ tile, using on-chip logic with no off-chip access.
This tile-local design preserves output regularity, introduces structured sparsity, and minimizes control and data movement overhead, making it naturally hardware-efficient.

We further adopt a block-wise scheduling strategy using sliding windows.  
Each block is processed in-stream, enabling local reuse and eliminating the need for global buffering.  
Our conflict-free memory layout (\Sec{subsec:layout}) supports parallel compression units without access contention, allowing \proj{} to scale with tile throughput at negligible latency.

In summary, \proj{} demonstrates effective {hardware-algorithm} co-optimization. 
Our vector-wise design improves redundancy detection and model fidelity, while streaming and tile-local execution ensure high hardware efficiency. 
This tightly integrated architecture makes \proj{} scalable, practical, and deployable for real-world VLM applications.

\section{\proj{} Architecture Overview}

\begin{figure}[t]
    \centering
    \includegraphics[width=\linewidth]{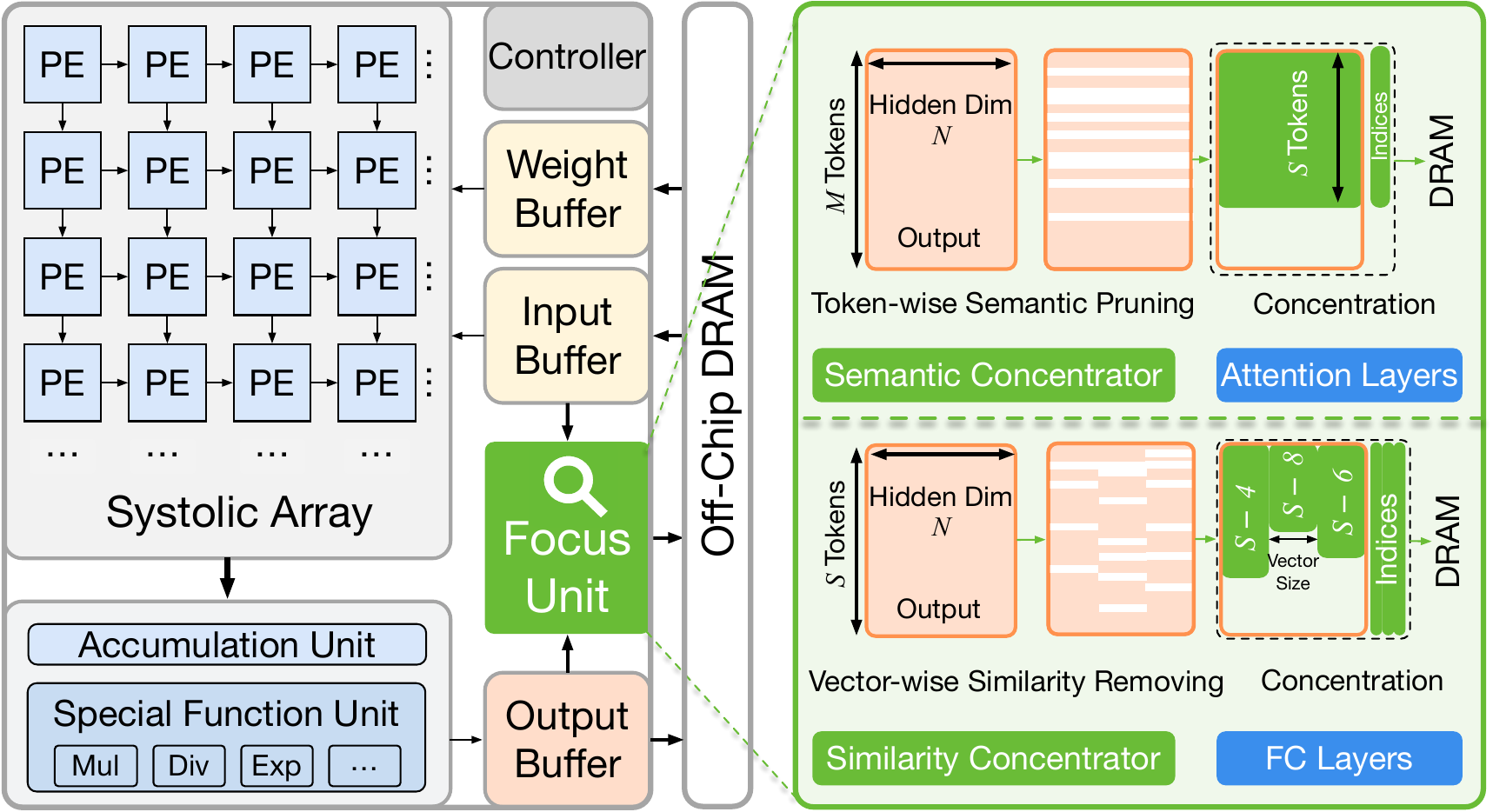}
    \caption{Overview of the \proj{} architecture. 
    The \proj{} Unit integrates a Semantic Concentrator (SEC) and a Similarity Concentrator (SIC), positioned between compute stages to eliminate redundancy before memory write-back. Both modules operate in a streaming manner and run entirely on-chip.
    }
    \label{fig:focus_arch}
\end{figure}

\proj{} introduces a modular \textbf{\proj{} Unit} to improve compute and memory efficiency in VLMs. 
As shown in \Fig{fig:focus_arch}, the \proj{} Unit is integrated near the memory interface of a standard systolic-array accelerator, intercepting data between compute stages without altering the core compute pipeline.
The \proj{} Unit consists of two streaming submodules:
\begin{itemize}
    \item \textbf{Semantic Concentrator (SEC)}: Performs token-level pruning in attention layers based on cross-modal Attention scores.
    \item \textbf{Similarity Concentrator (SIC)}: Performs vector-level redundancy elimination in fully connected (FC) layers, aligned with GEMM tiling.
\end{itemize}

\textbf{SEC} reduces the image token sequence length from $M$ to $S$.
It evaluates token importance using existing attention maps and prunes low-relevance tokens early in the pipeline.
Pruned tokens remain excluded in downstream layers, yielding cumulative savings in computation and memory access.

\textbf{SIC} further eliminates fine-grained redundancy among vectors within each GEMM tile.
It compares incoming vectors in a convolution-style window and replaces similar ones with index references to shared representatives.
This reduces the number of vectors processed per tile while preserving correctness via index-based reconstruction.

Both SEC and SIC operate entirely on-chip, support streaming dataflow, and dynamically adapt to data sparsity. 
By targeting complementary forms of redundancy from semantic and structural, \proj{} delivers efficient and scalable acceleration for Vision-Language Models.
We detail the hardware implementation of SEC and SIC in Sections~\ref{sec:sec} and~\ref{sec:sic}, respectively.
\section{Semantic Concentrator}
\label{sec:sec}

The \textbf{Semantic Concentrator (SEC)} enhances inference efficiency by selectively retaining semantically important visual tokens based on language context. 
It operates in the attention layers and consists of three tightly coordinated yet modular components, as shown in \Fig{fig:semantic}:
The \textbf{importance analyzer} that estimates the importance of visual tokens based on cross-modal attention.
A lightweight \textbf{top-$k$ sorter} that identifies the most important image tokens on the fly.
An \textbf{offset encoder} that enables lossless index tracking for streaming token recovery.

\subsection{Streaming Importance Analyzer}

\begin{figure}[t]
    \centering
    \includegraphics[width=\linewidth]{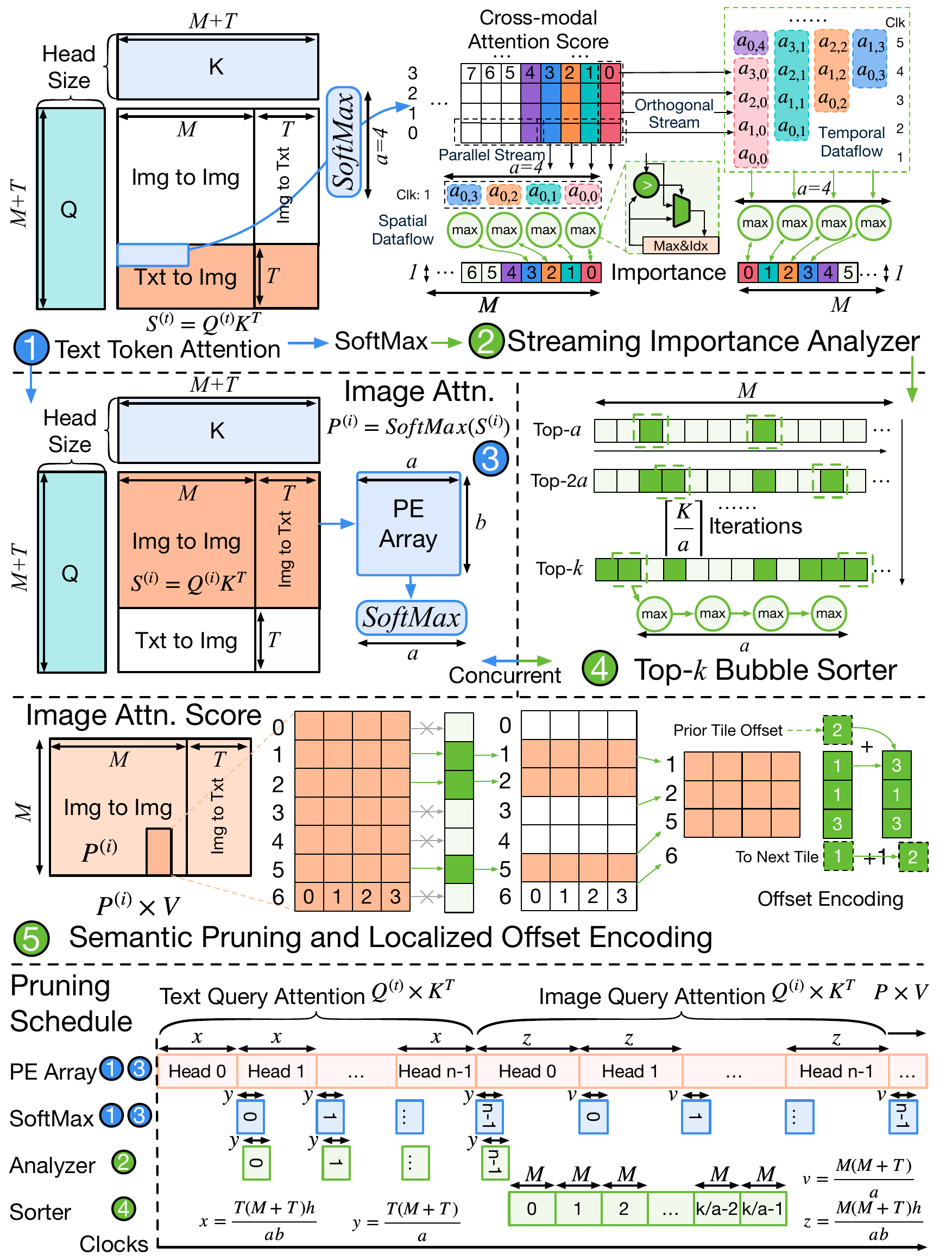}
    \caption{{Overview of the Semantic Concentrator (SEC), including the streaming importance analyzer, top-$k$ sorter, and offset encoder.}}
    \label{fig:semantic}    
\end{figure}

{The SEC integrates directly into the attention {$Softmax(QK^T)$} computation pipeline}. As shown in \Fig{fig:semantic}(1), for each attention head, it compute a {$Softmax(QK^T)$} matrix containing four blocks: image-to-image ($M \times M$), image-to-text, text-to-image ($T \times M$), and text-to-text. 
We extract the \textit{Text-to-Image} block ($T \times M$) as the cross-modal importance matrix $I$, where $M$ and $T$ represent the number of image and text tokens, respectively.
To estimate the importance of each image token $j$ over $n$ heads, we compute the maximum attention score it receives from any text token and all heads:
$
s_j = \max_{\substack{1 \leq k \leq n \\ 1 \leq i \leq T}} I^{(k)}_{i,j}
$.
This results in an importance vector of shape $1 \times M$ across all heads. {An on-chip buffer of 25 KB is used to store the importance vector.}

As depicted in \Fig{fig:semantic}(2), the importance analyzer uses {$a$ parallel max units to process the output of the attention \texttt{SoftMax} (provided by the special function unit).} To match throughput, {$a$ max units processes $a$ attention scores concurrently}. This streaming design supports two dataflows:
\textbf{Parallel (spatial) stream}: Attention columns are streamed directly into max units.
\textbf{Orthogonal (temporal) stream}: Attention rows are buffered locally, enabling column-wise reduction.

This fully streaming design ensures minimal area and latency overhead. Since no global operations are needed, the analyzer is decoupled from the main compute path and incurs negligible runtime cost.

\subsection{Top-$k$ Bubble Sorter}

Once the $1 \times M$ importance vector is computed, the system must identify the top-$k$ most relevant tokens. To avoid sorting all $M$ tokens globally, SEC adopts a pipelined bubble sorter as shown in \Fig{fig:semantic}(4).
By chaining the $a$ max units used earlier, we construct an $a$-way streaming bubble sorter. This structure incrementally refines the top-$a$ tokens, allowing us to compute top-$k$ selection over the $M$ candidates in $\frac{M \cdot k}{a}$ cycles, substantially more efficient than full sorting.

Crucially, this process is fully overlapped with the computation of image attention {($S^{(image)} = Q^{(image)}K^T$)}, which dominates the overall runtime.
Let the image attention (within {$QK^T$} GEMM) require $\frac{M \cdot (M+T) \cdot h \cdot n}{a \cdot b}$ cycles, where $h$ is the head dimension, $n$ is the number of heads, and $a \times b$ is the PE array size. The ratio of attention to the sorting operation is:
$
\frac{M \cdot (M+T) \cdot h \cdot n}{a \cdot b} \cdot \frac{a}{M \cdot k} = \frac{(M+T) \cdot h \cdot n}{k \cdot b}.
$
In typical configurations, $h \cdot n$ reaches into the thousands (e.g., 3584), while $b$ is much smaller (e.g., 32) and $k < (M + T)$. Therefore, the sorting operation completes well before the {$Q^{(i)}K^T$} finishes, ensuring that the SEC remains off the critical path and introduces no runtime bottleneck. {A scheduling diagram is shown in \Fig{fig:semantic} bottom to better understand the overlapping.}

\begin{figure*}[t]
    \centering
    \includegraphics[width=0.99\linewidth]{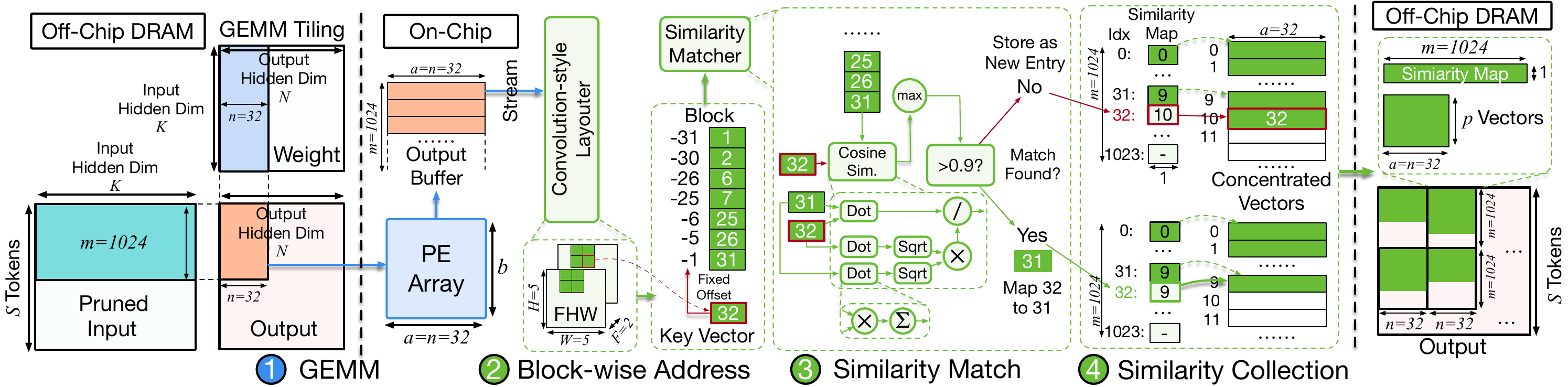}
    \caption{
    Overview of the Similarity Gather module. (1) GEMM tiling. (2) Convolution-style layout reorders outputs into a block-wise structure. (3) Cosine similarity is computed within blocks to detect and eliminate duplicates. (4) Only unique vectors are stored, while a similarity map enables reconstruction.
    }
    \label{fig:sic_gather}
\end{figure*} 

\subsection{Semantic Pruning and Offset Encoding}
\label{subsec:offset_encoding}
After identifying the top-$k$ most important tokens, we apply \textbf{semantic pruning} for the input of the subsequent attention operation, $P^{(i)}\times V$. 
As shown in \Fig{fig:semantic}(5), only the retained tokens are loaded and processed in $P^{(i)}\times V$, eliminating the need for any memory access or computation on pruned tokens.

In pure pruning mode, no additional metadata is needed. 
However, to support later stages (e.g., similarity concentration), we must record the position of retained tokens for spatial-temporal information. 
For this, the SEC generates \textbf{localized offset encodings}.
The offset encoder, shown in \Fig{fig:semantic}(5), operates in a sliding window over the retained tokens. 
For each token, it records a small integer representing its offset to the previous token. 
This compact encoding is sufficient to restore positional alignment for future operations (e.g., similarity matching).
The encoder's computation is local and streaming, requiring only lightweight registers and no global memory access.

Overall, the entire Semantic Concentrator, including the analyzer, sorter, and encoder, is fully modular and incurs minimal overhead.
SEC selectively retains the most informative visual tokens using cross-modal attention, top-$k$ selection, and compact position encoding. 
It operates transparently within the attention, requires no additional global memory access, and introduces negligible runtime or area overhead.

\section{Similarity Concentrator}
\label{sec:sic}

After semantic-level token pruning, subsequent FC layers operate seamlessly on the reduced token set, as the pruning is token-aligned and preserves structural layout. However, semantic pruning alone does not address fine-grained redundancy at the vector level. In contrast, \textbf{Similarity Concentration} targets vector-wise redundancy by matching and merging similar output vectors within local regions across frames.

Different from semantic pruning, where unimportant tokens are discarded, similar vectors often carry essential information and thus require accurate removal and later reconstruction. To support this, the similarity process includes two core components:
\textbf{Similarity Gather}: removes similar vectors and constructs a compact output.
\textbf{Similarity Scatter}: restores the original full layout using a similarity map.

\subsection{{Similarity Gather}}
\label{subsec:gather}
In this section, we first detail the design of \textbf{Similarity Gather}, which efficiently operates in a fully streaming fashion.

\textbf{GEMM Tiling.}  
As shown in \Fig{fig:sic_gather}(1), Similarity Gather operates on the output of GEMM\footnote{{Similarity Gather on output of FFN, O projection, and PV GEMM}}. Assume the input has dimension $M \times K$, the weight matrix is $K \times N$, and the output is $M \times N$. 
Due to limited on-chip resources, we adopt a tiling strategy widely used in modern accelerators~\cite{chen2016eyeriss,jouppi2017datacenter}. 
Specifically, the input and weight tiles are of size $m \times K$ and $K \times n$, and the output tile is $m \times n$, where we typically set $m = 1024$ and $n = 32$.
The PE array performs one output tile at a time, producing $a$-length output vectors, where $a = n = 32$ in our implementation. {These vectors are then streamed to the Similarity Gather logic for processing when an output tile gets ready}.

\begin{figure*}[t]
    \centering
    \includegraphics[width=\linewidth]{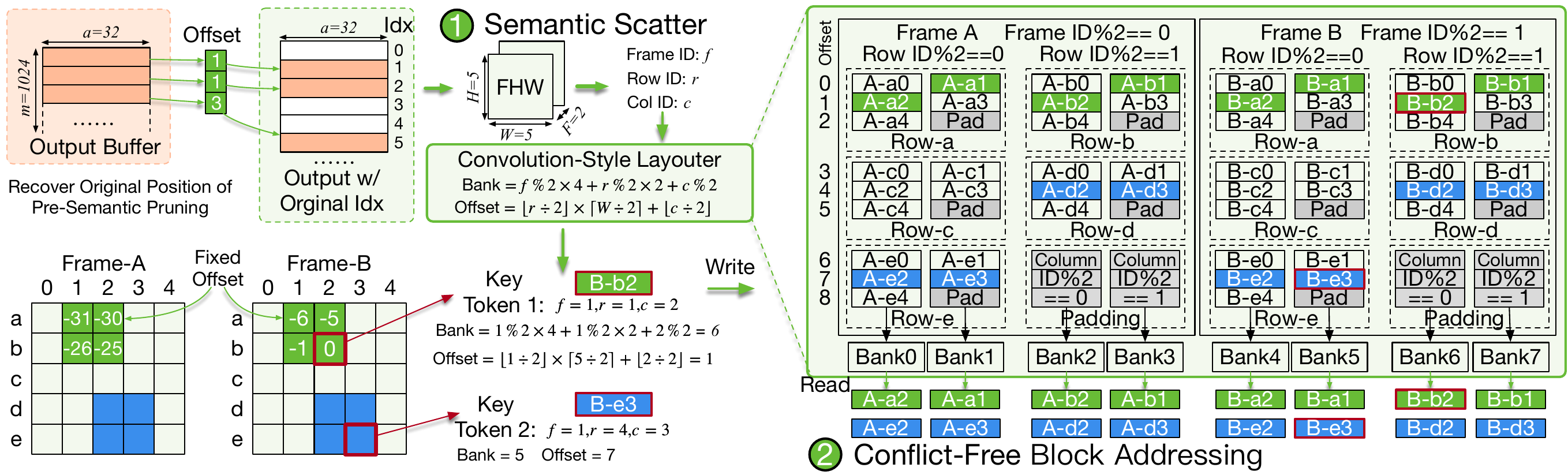}
    \caption{The convolution-style layouter enables accurate token positioning and conflict-free memory access for block-level similarity. (1) Reconstructs token positions via semantic offsets and maps outputs to a FHW-layout. (2) Enables conflict-free addressing across memory banks without data duplication.}
    \label{fig:sic_layouter}
\end{figure*}

\textbf{Block-wise Addressing.}  
To exploit spatiotemporal redundancy, we adopt a \textbf{convolution-style layout} over two adjacent frames, as illustrated in \Fig{fig:sic_gather}(2). 
{A $2 \times 2 \times 2$ sliding window spans both spatial and temporal dimensions with a stride of 1, forming a block that contains 8 vectors, 4 from each frame. 
Each element in the block is an $a$-dimensional output vector produced by the GEMM operation, and the block serves as a localized comparison group for redundancy detection.}

To efficiently support this structure, GEMM outputs are dynamically reorganized into the convolution-style layout using a dedicated reordering module, as detailed in \Sec{subsec:layout}. 
Within each block, the vector with the highest index (e.g., token ID 32) is selected as the key vector and compared against the other 7 vectors in the same block (e.g., token IDs 1 through 31) to identify potential redundancy.

\textbf{Vector-wise Similarity Matching.}  
As shown in \Fig{fig:sic_gather}(3), each key vector is streamed into the \textbf{Similarity Matcher}, which performs localized comparisons to determine whether the vector is redundant. 
We adopt cosine similarity to compare two vectors $\mathbf{p}$ and $\mathbf{q}$ of length 32:
\vspace{-3mm}
\[
\frac{\mathbf{p} \cdot \mathbf{q}}{\|\mathbf{p}\| \cdot \|\mathbf{q}\|} = 
\frac{\sum_{i=1}^{32} p_i q_i}{
\sqrt{\sum_{i=1}^{32} p_i^2} \cdot 
\sqrt{\sum_{i=1}^{32} q_i^2}}.
\]

Thanks to the regularity of the convolution-style layout, each token can precompute its L2-norm ($\|\mathbf{p}\|$) and store it in a buffer. 
This allows the matcher to perform similarity comparisons using only a single dot-product unit and a small number of low-overhead element-wise operations.
Furthermore, the vector-level granularity reduces the normalization length to 32, greatly simplifying the hardware design compared to token-wise similarity matching. 
In contrast, prior accelerators such as AdapTiV~\cite{yoo2024adaptiv} and CMC~\cite{song2024cmc} compute similarity at the token level, requiring expensive global memory access and full-sequence comparisons. 
By operating at the vector level within localized blocks, \proj{} achieves significantly lower matching overhead while maintaining semantic fidelity.

In practice, most of these operations are already supported by the Special Function Unit (SFU), which is commonly used for \textit{RMSNorm}~\cite{zhang2019root} and \textit{SoftMax} computations.
Compared to these more complex operations, cosine similarity is lightweight and well-suited for hardware acceleration. 
Although the matcher could reuse existing SFU logic, for fairness, we implement it as a separate module and include its area and energy in our evaluation.
The total overhead remains minimal, accounting for {<1\%} of the systolic array design.

It is worth noting that similarity matching is \textit{not} on the critical path of GEMM, as comparisons are performed only once per output tile.  
For a tile with $m = 1024$ vectors, each requiring 7 pairwise comparisons and 1 L2-norm computation (based on the $2 \times 2 \times 2$ block structure), the matcher needs at most $8 \times m$ cycles to process the tile.  
In contrast, GEMM requires $\frac{K}{b} \times m$ cycles, where $K$ is the hidden dimension and $b$ is the number of PE rows.  
In our setup, with $K = 3584$ and $b = 32$, GEMM takes $112 \times m$ cycles per tile, far exceeding the cost of similarity matching.  
Only when $K < 256$ does the matcher approach the critical path.

To address this corner case, we can scale the design by deploying multiple matcher units in parallel. 
Our convolution-style layout inherently supports conflict-free parallel access, allowing similarity matching to be fully overlapped with GEMM computation without introducing additional latency.

\textbf{Similarity {Collection}.}  
Once similarity matching completes, each vector has two outcomes:  
{No match:} The vector is unique and added to the concentrated output buffer.  
{Match found:} The vector matches a previously stored one (e.g., token 32 matches token 31), and we reuse the index of the matched token.

To support lossless reconstruction, we maintain a \textbf{Similarity Map} of size $1 \times m$ per tile. 
This map records, for each of the original $m$ output vectors, the index of its representative in the compact buffer. 
For instance, if token 32 matches token 31, we assign token 32 the index “9” from token 31.
After processing all $m = 1024$ vectors in a tile, only the deduplicated vectors and the similarity map are written back to DRAM. 
This significantly reduces memory bandwidth and storage.

All stages of this pipeline, including reordering, matching, and mapping, are performed on-chip, in a streaming fashion, without global synchronization or off-chip overhead. 
This localized similarity removal aligns naturally with GEMM tiling and preserves high data locality throughout execution.

\subsection{Convolution-style Layouter}
\label{subsec:layout}
We now describe the design of the \textit{convolution-style layouter}, which addresses two key challenges in enabling efficient block-level similarity matching after semantic pruning: 
(1) recovering token positions and (2) avoiding memory access conflicts during parallel execution.

\textbf{Challenge 1: Recovering Token Positions after Pruning.}  
Semantic pruning disrupts the spatial structure of tokens by removing unimportant entries, making it nontrivial to identify the 2D position of retained tokens in the original frame. 
To enable meaningful $2 \times 2 \times 2$ comparisons across adjacent frames, we must reconstruct each token’s (Frame, Height, Width) coordinate after pruning.

As shown in \Fig{fig:sic_layouter}(1), we achieve this using the \textit{offset encoding} generated during the semantic pruning stage (see \Sec{subsec:offset_encoding}). 
This offset, streamed alongside the GEMM output, allows us to recover the original spatial location of each token. 
Tokens are then reorganized into a structured 3D tensor layout following the FHW (Frame–Height–Width) order to support localized block grouping.

\textbf{Challenge 2: Avoiding Memory Conflicts in Parallel Matching.}  
To form a $2 \times 2 \times 2$ spatiotemporal block, vectors are drawn from multiple rows, columns, and frames. 
A naive layout may introduce bank conflicts or require data duplication across SRAM banks, an approach used by traditional CNN accelerators~\cite{chen2016eyeriss} but with significant memory overhead (up to 8$\times$ replication).

\begin{figure*}[t]
    \centering
    \includegraphics[width=0.98\linewidth]{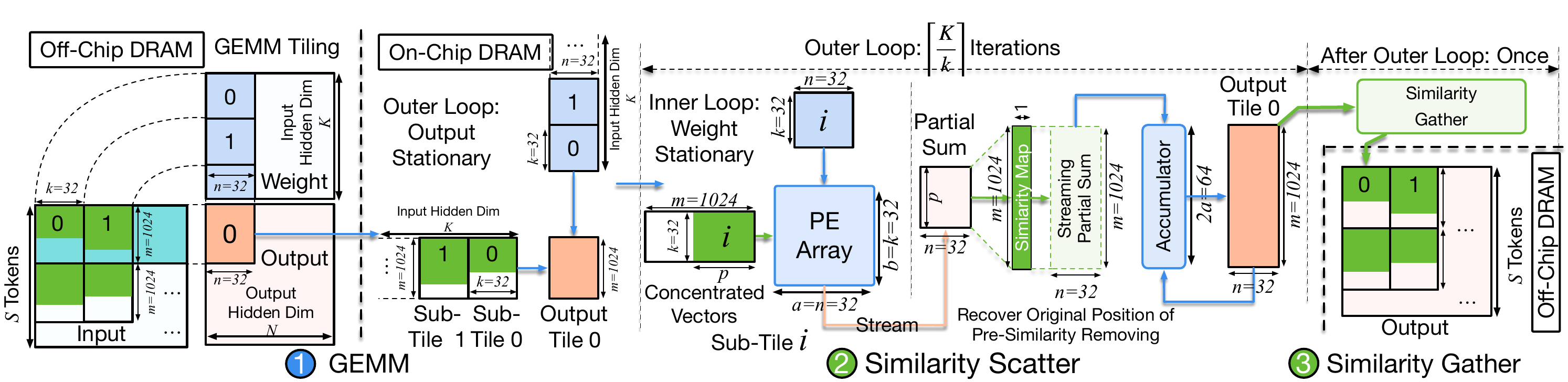}
    \caption{GEMM tiling and Similarity Scatter design. (1) GEMM computes over concentrated vectors. (2) Similarity Scatter reconstructs and accumulates vector results using similarity maps. (3) Final output is passed to Similarity Gather once after all iterations in a tile.}
    \label{fig:sic_scatter}
\end{figure*}

To eliminate these conflicts, we propose a \textbf{conflict-free convolution-style layout}, shown in \Fig{fig:sic_layouter}(2), which deterministically maps each token to a unique bank and offset based on its FHW position. 
Given frame index $f$, row $r$, and column $c$, the memory bank and address are computed as:
\vspace{-1mm}
\[
\texttt{Bank} = f \bmod 2 \times 4 + r \bmod 2 \times 2 + c \bmod 2,
\]
\vspace{-3mm}
\[
\texttt{Offset} = \left\lfloor \frac{r}{2} \right\rfloor \times \left\lceil \frac{W}{2} \right\rceil + \left\lfloor \frac{c}{2} \right\rfloor,
\]
where $W$ is the width of the frame. 
This mapping guarantees that all 8 vectors in any $2 \times 2 \times 2$ block reside in distinct memory banks and can be read simultaneously without contention.

\textbf{Key Insight:}  
Unlike traditional approaches that duplicate inputs to avoid access conflicts, our layout achieves fully parallel, conflict-free access \textit{without any data replication}. 
This enables streaming similarity matchers to operate in parallel across tiles and spatial regions, scaling throughput without modifying the GEMM pipeline. 
The layouter thus plays a critical role in supporting parallel similarity execution and maintaining high utilization.

\subsection{Similarity Scatter}
\label{subsec:scatter}

\textbf{GEMM Tiling for Concentrated Vectors.}
As shown in \Fig{fig:sic_scatter}(1), Similarity Scatter operates on the concentrated vectors generated from earlier stages. 
Since only a subset of the original $m=1024$ tokens is retained in each tile ($p < 1024$), the input to this GEMM stage is \textbf{logically sparse} but \textbf{structurally dense}. 
To maintain compatibility with standard systolic-array architectures, GEMM is performed using a conventional tiling scheme with dimensions $m=1024$ and $n=32$.
The GEMM execution follows a two-level nested loop structure:
The \textbf{outer loop} adopts an \textit{output-stationary} dataflow, keeping the $m \times n$ output tile resident on-chip to accumulate results across the $K$ dimension.
The \textbf{inner loop} follows a \textit{weight-stationary} strategy, loading one $k \times n$ weight sub-tile into the PE array while streaming in a $p \times k$ sub-tile of concentrated input vectors.

Each inner loop iteration computes partial products and generates one $a$-dimensional partial sum vector per cycle. Our vector size 32 matches with the $k$ tile size and array height $b$, ensuring full utilization of the PE array.
These vectors are streamed out and accumulated over successive iterations to form the final tile result.
The key advantage arises from the reduced number of active input vectors ($p < 1024$), which significantly lowers the computational workload. 
However, since different sub-tiles may have different subsets of concentrated vectors, each possibly representing multiple original tokens, direct accumulation would produce incorrect outputs due to semantic aliasing.

\textbf{Similarity Scatter and Gather.}
To resolve this, we introduce the \textbf{Similarity Scatter} module, illustrated in \Fig{fig:sic_scatter}(2). 
After each GEMM step, the generated partial sums are streamed into a temporary buffer. 
Using the similarity map from previous layer's the gather phase (see \Sec{subsec:gather}), each partial sum is replicated and redistributed to its associated original token indices, reconstructing the full $m=1024$ output.
This scattered output is then accumulated into an output-stationary buffer spanning all outer loop iterations. 
To maintain throughput parity, we employ a $2a$-wide accumulator (e.g., 64 when $a = 32$), enabling concurrent accumulation of reconstructed vectors and streaming outputs. 
The reconstruction process is performed in-place, incurs negligible overhead, and does not require additional memory allocation.

Upon completing all $\left\lceil \frac{K}{k} \right\rceil$ outer loop iterations, the fully accumulated output tile is passed to the \textbf{Similarity Gather} unit (see \Sec{subsec:gather}), shown in \Fig{fig:sic_scatter}(3). 
This final stage is invoked only once per tile after GEMM concludes and lies entirely off the critical compute path.

In summary, by executing GEMM on a compact set of concentrated vectors, \proj{} achieves substantial compute savings.  
Through the Similarity Scatter module, it efficiently reconstructs full output tiles with minimal accuracy loss, and the final gather stage removes vector-level redundancy.  
This hardware-oriented, vector-granular compression strategy ensures high compute efficiency while preserving model fidelity.  
Our evaluation shows that the additional logic is lightweight and does not impact GEMM throughput, making it a key enabler of \proj{}'s performance advantage.

\section{Evaluation}
\label{sec:eval}
\subsection{Methodology}

\textbf{Evaluation Models and Datasets.}
We evaluate \proj using three representative VLMs with video understanding and reasoning capabilities: Llava-OneVision-7B (Llava-OV)~\cite{li2024llava}, Llava-Video-7B (Llava-Vid)\cite{zhang2024video}, and MiniCPMV-2.6 (MiniCPM)~\cite{yao2024minicpm}. These models are tested on three widely adopted video understanding benchmarks: VideoMME (VMME)~\cite{fu2025video}, MVBench (MVB)~\cite{li2024mvbench}, and MLVU~\cite{zhou2025mlvu}. These datasets include diverse video types and durations, enabling a holistic evaluation of model capabilities across comprehension, temporal reasoning, and multimodal alignment.
We use open-source models obtained from HuggingFace Transformers~\cite{wolf2020transformers} and perform evaluation via the \texttt{lmms-eval}~\cite{zhang2024lmmsevalrealitycheckevaluation} multimodal benchmarking framework to ensure consistency and fairness.

\textbf{Baselines.}
We compare \proj against two state-of-the-art architectures: AdapTiV~\cite{yoo2024adaptiv}, a vision transformer accelerator, and CMC~\cite{song2024cmc}, an accelerator optimized for video transformers. We extend their designs to make them compatible with VLMs. {CMC performs inter-frame similarity checks, whereas AdapTiV focuses on intra-frame similarity detection; both exclude text tokens.} We also compare with the vanilla systolic array~\cite{kung1979systolic} architecture for a base reference.
In addition to hardware baselines, we also compare with FrameFusion~\cite{fu2024framefusion}, a state-of-the-art token pruning algorithm tailored for efficient VLMs with video inputs. 

\begin{table}[b]
    \centering
    \caption{{\proj Architecture Setup}}
    \resizebox{0.5\textwidth}{!}{
    \begin{tabular}{l|l}
    \toprule
        PE Array & 32 $\times$ 32; FP16 Mul FP32 Acc; Weight Stationary \\ \midrule
        \proj Hyper-params & Block Size: 2$\times$2$\times$2; Vector Length: 32; \\
        & Similarity Threshold: 0.9; M Tile Size: 1024\\
        & {Semantic: Retain 40\%/30\%/20\%/15\%/10\%} \\
        & {of total image tokens at layer 3/6/9/18/26} \\ \midrule
        On-Chip Buffer & Input: 128KB; Weight: 78KB; Output: 512KB; \\
        & Layouter Buffer: 16KB for 256-vector window; \\
        & 734KB in total.
                      \\ \midrule
        Off-Chip Memory & {DDR4 4Gb × 16, 2133R, 4 Channels, 64GB/s} \\ 
    \bottomrule
    \end{tabular}
    }
    \label{tab:hardware-config}
\end{table}
\textbf{Algorithm Implementation of \proj and Baselines.}
We implement the algorithm of our proposed \proj method in PyTorch~\cite{paszke2019pytorch}. For the baselines, we faithfully reproduce the token pruning algorithm from AdapTiV and CMC, carefully tuning their hyperparameters for application to VLMs. For FrameFusion, we adopt the official open-source implementation without modification. All algorithms are executed on an NVIDIA A100 GPU~\cite{nvidia2020a100} using FP16 precision for fair and consistent comparison.

\textbf{Architecture Implementation of \proj and Baselines.}
Our \proj architecture setup is shown in \Tbl{tab:hardware-config}. To evaluate architectural performance, we develop a cycle-accurate simulation framework based on SCALEsim-v2~\cite{samajdar2018scale}. {The simulator accepts layer-wise sparse traces generated from specific models and datasets in our PyTorch implementation, enabling precise modeling of cycles and memory access.} {We implement the \proj architecture in SystemVerilog and generate the on-chip SRAMs using the TSMC N28HPC+ Memory Compiler. The RTL is synthesized with a target clock period of 1.32 ns ($\approx$757 MHz) under the worst-case slow–slow (SS) corner (0.81V, 125°C), achieving 0 ns worst negative slack (WNS) and providing a 34\% timing margin for place-and-route at 500 MHz. The resulting area is reported from post-synthesis analysis, and the on-chip power is obtained from post-synthesis simulation using Synopsys Design Compiler. Off-chip DRAM energy is modeled with DRAMsim3~\cite{li2020dramsim3} for device-level power.}
For a fair comparison, we also implement the core logic of all baseline accelerators in SystemVerilog and evaluate their area and energy using the same toolchain as \proj.

\begin{table}[t]
    \centering
    \caption{Accuracy and Computation Sparsity of \proj and Baselines}
    \resizebox{0.47\textwidth}{!}{
    \begin{tabular}{l|l|l|c|c|c|c|c}
    \toprule
        Models & Dataset & Metric & Ori. & FF & Ada. & CMC & Ours \\ \midrule
        \multirow{6}{*}{Llava-Vid} 
        & \multirow{2}{*}{VMME} & Acc. & 64.15 & 62.00 & 62.44 & 62.52 & \textbf{62.74} \\ 
        & ~ & Sparsity & 00.00 & 70.00 & 52.15 & 58.62 & \textbf{82.82} \\ \cmidrule{2-8}
        & \multirow{2}{*}{MLVU} & Acc. & 67.74 & 65.38 & 65.94 & 65.17 & \textbf{65.99} \\ 
        & ~ & Sparsity & 00.00 & 70.00 & 32.52 & 42.46 & \textbf{78.26} \\ \cmidrule{2-8}
        & \multirow{2}{*}{MVB} & Acc.  & 60.33 & 57.20 & 57.73 & 58.18 & \textbf{58.20} \\ 
        & ~ & Sparsity & 00.00 & 70.00 & 41.07 & 53.00 & \textbf{78.44} \\ \midrule
        \multirow{6}{*}{Llava-OV} 
        & \multirow{2}{*}{VMME} & Acc. & 58.41 & 57.70 & 58.33 & 58.11 & \textbf{58.70} \\ 
        & ~ & Sparsity & 00.00 & 70.00 & 36.80 & 47.95 & \textbf{81.49} \\ \cmidrule{2-8}
        & \multirow{2}{*}{MLVU} & Acc. & 63.32 & \textbf{62.54} & 62.22 & 62.50 & 62.52 \\ 
        & ~ & Sparsity & 00.00 & 70.00 & 39.55 & 35.48 & \textbf{78.34} \\ \cmidrule{2-8}
        & \multirow{2}{*}{MVB} & Acc.  & 58.38 & \textbf{56.93} & 56.83 & 56.75 & 56.78 \\ 
        & ~ & Sparsity & 00.00 & 70.00 & 42.03 & 63.69 & \textbf{85.49} \\ \midrule
        \multirow{6}{*}{MiniCPM} 
        & \multirow{2}{*}{VMME} & Acc. & 58.81 & \textbf{58.81} & 58.07 & 55.89 & 58.30 \\ 
        & ~ & Sparsity & 00.00 & 70.00 & 49.27 & 57.20 & \textbf{82.87} \\ \cmidrule{2-8}
        & \multirow{2}{*}{MLVU} & Acc. & 55.89 & 54.80 & \textbf{54.84} & 43.80 & 53.59 \\ 
        & ~ & Sparsity & 00.00 & 70.00 & 41.88 & 35.23 & \textbf{78.01} \\ \cmidrule{2-8}
        & \multirow{2}{*}{MVB} & Acc.  & 55.63 & 52.43 & 53.70 & 48.78 & \textbf{54.30} \\ 
        & ~ & Sparsity & 00.00 & 70.00 & 50.09 & 40.27 & \textbf{75.99} \\ 
    \bottomrule
    \end{tabular}
    \label{tab:accuracy}
    }    
\end{table}

\subsection{Algorithmic Accuracy and Theoretical Sparsity}

To evaluate the effectiveness of the multilevel concentration technique in \proj, we compare both model accuracy and the achieved computational sparsity against baseline methods. The computation sparsity is calculated through the ratio of the number of operations using the method to the number of operations required by the systolic array with original input. The results are presented in \Tbl{tab:accuracy}.

\proj consistently achieves the highest accuracy across most evaluated scenarios, outperforming both software-only methods and hardware-based approaches. Compared to the original, uncompressed models, the average accuracy degradation with \proj is only 1.20\%, demonstrating its ability to preserve semantic fidelity.

In addition to maintaining high accuracy, \proj also achieves the highest computational sparsity across all models and datasets. Specifically, \proj achieve sparsity of \textbf{80.19$\%$} on average, delivering 37.37\% and 31.98\% higher sparsity than AdapTiV and CMC, respectively, and outperforms FrameFusion by over 10.19\% in sparsity.

\begin{figure*}[t]
    \centering
    \includegraphics[width=1.0\linewidth]{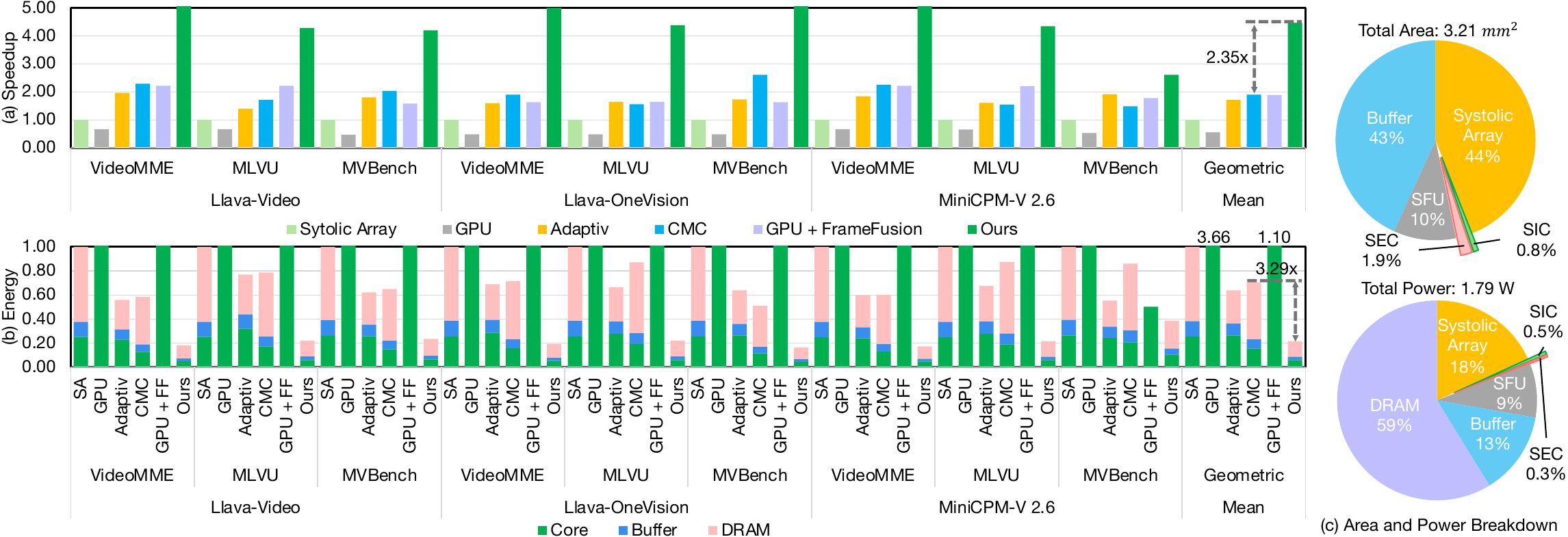}
    
    \caption{{Left: Speedup and Energy efficiency.} Right: Area and power breakdown}
    \label{fig:main}
\end{figure*}

\subsection{Performance and Energy Evaluation}

We compare \proj against baseline methods, including the vanilla systolic array (SA), Adaptiv, and CMC, across multiple VLM models and datasets. The architectural setup for all baselines and \proj is detailed in \Tbl{tab:setup_compare}. We maintain the same frequency, technology node, number of processing elements, operand bit width, and DRAM bandwidth across all designs. {We further compare against the performance on an NVIDIA Jetson Orin Nano GPU~\cite{nvidia_jetson_orin_nano_2023}, evaluated with and without the FrameFusion algorithm.} The performance and energy results are presented in \Fig{fig:main} (a) and (b).

\textbf{Performance.}  
\proj achieves significant performance improvements across all benchmarks. On average, it delivers a 4.47$\times$ speedup over the vanilla systolic array, which process dense input. This improvement stems from the ability of multi-level concentration to aggressively compress input tokens, getting 80.2\% sparsity. 

Compared to AdapTiV, \proj achieves a \textbf{2.60$\times$} average speedup. While AdapTiV effectively detects and prunes nearby redundant visual tokens, it operates at a coarser granularity. In contrast, \proj performs vector-wise similarity removal, enabling finer-grained redundancy elimination. 

Against CMC, \proj achieves a \textbf{2.35$\times$} speedup. While CMC leverages external video codecs to perform wide-range redundancy search, this approach is often inefficient due to a high rate of mismatches. In contrast, \proj efficiently identifies sufficient redundancy within localized blocks using its on-chip similarity matcher.

{Compared with the GPU, our design achieves a $7.90\times$ speedup over the GPU and a $2.37\times$ speedup over the GPU running with FrameFusion. This improvement stems from our architecture’s ability to achieve higher computational utilization than the GPU. Moreover, \proj attains higher sparsity than FrameFusion due to its finer-grained redundancy removal, which is difficult to exploit on GPU Tensor Cores. 
}

\textbf{Energy Efficiency.}
As shown in \Fig{fig:main}(b), we report the total energy consumption of \proj and baseline designs, normalized to the vanilla systolic array (SA). The energy breakdown includes three components: on-chip core, on-chip buffer, and off-chip memory.

Compared to SA, {GPU,} AdapTiV, CMC, and {GPU with FrameFusion}, \proj achieves average energy efficiency improvements of {\textbf{4.67$\times$}}, {\textbf{17.09$\times$}} {\textbf{2.98$\times$}}, {\textbf{3.29$\times$}}, and {\textbf{5.13$\times$}}, respectively. These results highlight that \proj delivers significant savings across both computation and memory under constrained on-chip budget. This efficiency gain stems from our architecture's ability to sparsify output of GEMM on-chip immediately after output tile generation, ensuring that all subsequent off-chip memory transactions operate on compressed data. A detailed analysis of memory access reduction is presented in \Sec{sec:mem_access}.

\begin{table}[t]
    \centering
    \caption{Configuration comparison of \proj and baseline architecture}
    \resizebox{0.5\textwidth}{!}{
    \begin{tabular}{l|c|c|c|c}
    \toprule
        Architecture & SystolicArray & Adaptiv & CMC & Ours \\ \midrule
        Technology & 28nm & 28nm & 28nm & 28nm \\ \midrule
        Frequency & 500MHz & 500MHz & 500MHz & 500MHz \\ \midrule
        \multirow{2}{*}{PE Array} & 32x32 & 16x64 & 32x32 & 32x32 \\ 
         & 16-bit & 16-bit & 16-bit & 16-bit \\ \midrule
        Buffer Size& 734KB & 768KB & 907KB & 734KB \\ \midrule
        DRAM Bandwidth & 64GB/s & 64GB/s & 64GB/s & 64GB/s \\ \midrule
        On-chip Area/$mm^2$ & {3.12} & {3.38} & {3.58} & {3.21} \\ \midrule
        On-chip Power/mW & {720} & {1176} & {832} & {736} \\ \bottomrule
    \end{tabular}
    }
    \label{tab:setup_compare}
\end{table}

\textbf{Area and Power Analysis.}
The area and power consumption of \proj and the baselines are also summarized in \Tbl{tab:setup_compare}. The power statistics is derived on Llava-Video-7B with VideoMME dataset. Our \proj design occupies {3.21}\,mm\textsuperscript{2} of on-chip area and consumes {736}\,mW of power, both of which are lower than those of Adaptiv and CMC.

\proj is smaller than CMC, as the external video codec used by CMC incurs substantial hardware overhead. Compared to Adaptiv, which adopts a lightweight similarity detection mechanism, \proj remains more efficient due to its streaming SEC that operates on localized input. Despite its enhanced functionality, \proj introduces only a {\textbf{2.7\%}} increase in area and a {\textbf{0.9\%}} increase in power consumption relative to the systolic array architecture.
These results highlight the efficiency and low overhead of the \proj unit, which delivers significant performance benefits within a modest hardware budget.

To gain a deeper understanding of the overhead introduced by \proj, we present a detailed breakdown in \Fig{fig:main}(c). We observe that the proposed Semantic Concentrator and Similarity Concentrator are both highly lightweight, accounting for only 1.9\% and 0.8\% of the overall area, respectively. These two units also contribute negligibly to the overall power consumption. This demonstrates that SEC and SIC are well-suited for resource-constrained scenarios.

{
\textbf{Overall Insights.} 
Beyond speedup and energy gains, \proj{} establishes a new paradigm for redundancy-aware VLM acceleration through tight algorithm–hardware co-design. 
At the \textbf{token level}, it performs on-the-fly \textit{Top-$k$ detection} via streaming processing, handling sparsity in real time with minimal cost.  
At the \textbf{block level}, a \textit{block-wise sliding window} propagates local similarity using only on-chip resources, reducing memory and buffer demand.  
At the \textbf{vector level}, \proj{} applies \textit{vector-wise similarity pruning} with a \textit{gather–scatter} scheme to control fine-grained irregular access and fully exploit sparsity.  
Together, these techniques translate algorithmic sparsity into tangible performance gains with minor hardware complexity.
}

\begin{figure}
    \centering
    \includegraphics[width=1\linewidth]{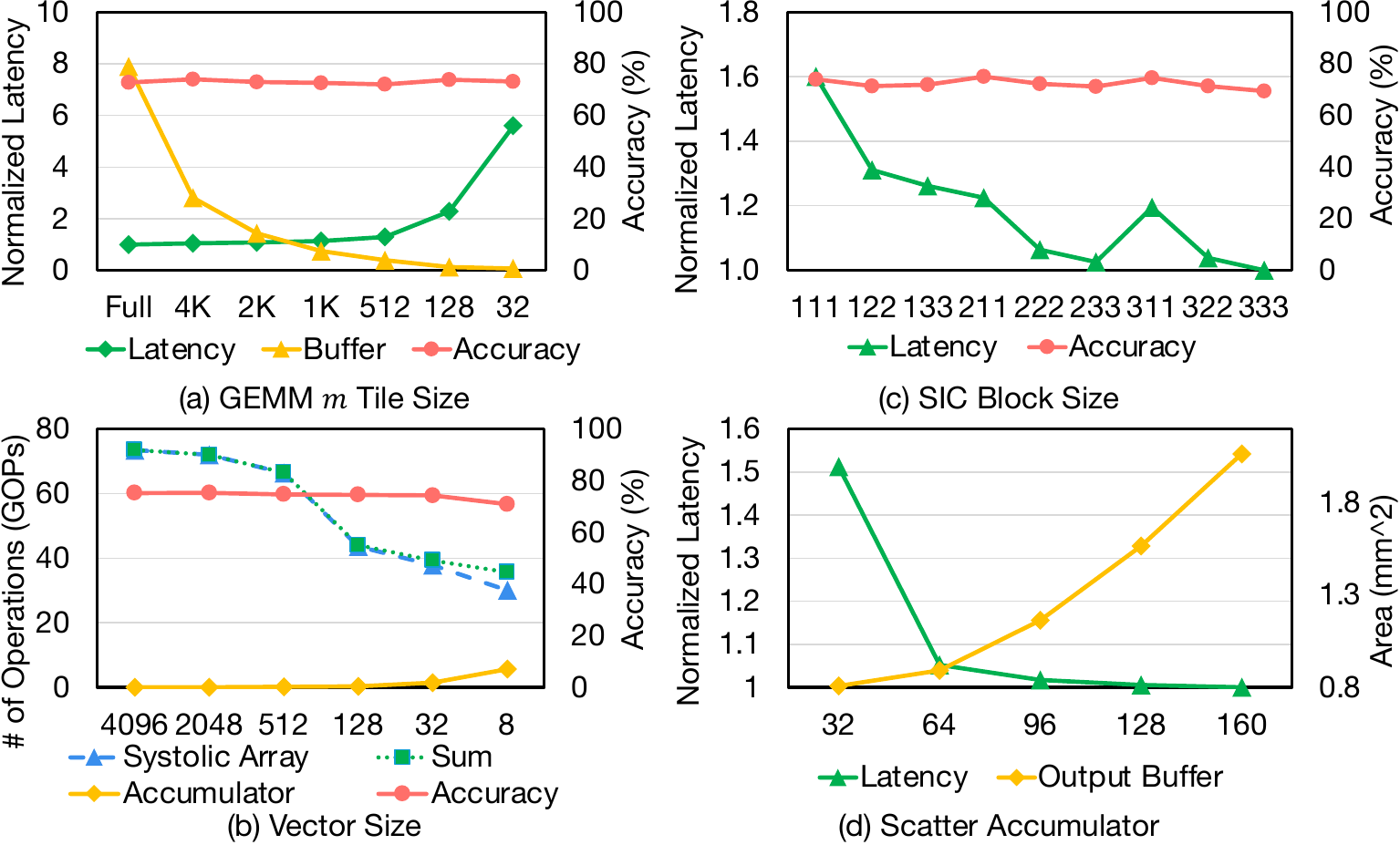}
    \caption{{Design Space Exploration}}
    \label{fig:DSE}
\end{figure}

\subsection{Design Space Exploration}

To evaluate the impact of key architectural parameters in \proj{}, we conduct a comprehensive design space exploration. We focus on four primary factors, varying each individually while fixing the others to their default values to isolate their effects. 
{Note that architectural parameters, other than the number of scatter accumulators, may also affect model accuracy. We evaluate accuracy under these variations and observe that the impact is generally negligible, allowing us to safely prioritize performance in our design exploration.}
All measurements are taken on the Llava-Video-7B model, using either the VideoMME or MLVU dataset.

\textbf{GEMM $m$ Tile Size.}
{As shown in \Fig{fig:DSE}(a), we sweep the tile size from the full input height down to 32. As the tile size decreases, the end-to-end latency steadily increases. {This trend arises because similarity gathering operates per tile. When a 2×2×2 block crosses tile boundaries, Focus only compares tokens within the same tile as the key token. For example, when the first token of a tile is the key, its neighbors outside the tile are unavailable for comparison. With smaller tile sizes (e.g., $m=32$), such boundary-crossing cases become more frequent, causing potentially similar vectors to be treated as distinct due to the limited comparison scope.}

While larger tiles offer better compression, they require more on-chip buffer to store intermediate results, increasing area and power consumption. We observe a trade-off between latency and buffer usage. From the latency–buffer curve in \Fig{fig:DSE}(a), a tile size of 1024 emerges as an optimal design point. It incurs only 19\% higher latency compared to the full-height tile while substantially reducing buffer requirements to a practical level.

\textbf{Vector Size.}
Vector size determines the granularity of similarity concentration and directly impacts the sparsity and operation counts. To assess this, we measure the number of operations of a layer in two main components of \proj{}: (1) MAC operations in the main systolic array, and (2) accumulation operations in the outer accumulator during Similarity Scatter.

As shown in \Fig{fig:DSE}(b), reducing the vector size leads to fewer operations in the systolic array. This is because smaller vectors enable finer-grained similarity comparisons, allowing more aggressive redundancy removal and reducing the input size to the PE array. However, smaller vector sizes also increase the number of K-dimension iterations, requiring more frequent accumulation, which in turn raises the operation count in the accumulator.

Beyond operation count, the systolic array dimension $b$ must be equal to or less than the vector size to utilize the benefits of fine-grained input. Taking both operational efficiency and hardware compatibility into account, we identify a vector size of 32 as an optimal design point, achieving strong compression while maintaining high utilization of the systolic array.

\begin{figure}[t]
    \centering
    \includegraphics[width=0.99\linewidth]{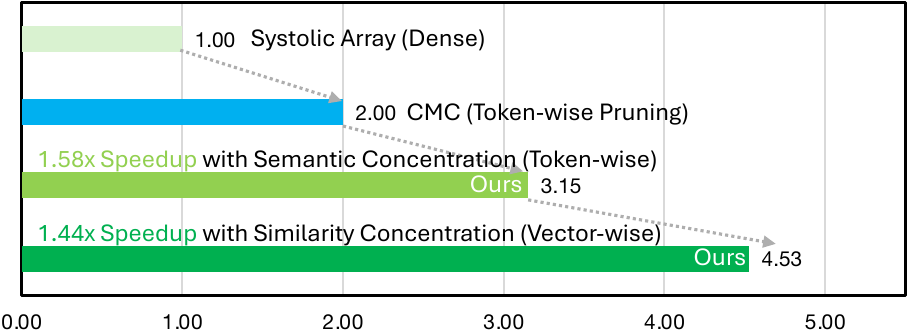}    
    \caption{Ablation Study for \proj}
    \label{fig:ablation}    
\end{figure}

\textbf{Similarity Concentrator Block Size.}
The block size used in the SIC directly impacts the spatial and temporal context available for similarity detection. 
{We vary the block size along both the temporal (frame) and spatial (height and width) dimensions to examine its impact on performance.}
{As shown in \Fig{fig:DSE}(c), the three-digit labels on the x-axis denote block sizes across these dimensions (e.g., 122 indicates f=1,h=2,w=2). We observe that enlarging the block size in either temporal or spatial dimensions reduces latency, as larger blocks provide broader context for similarity detection.}
{Notably, extending the block size along the temporal dimension yields a more pronounced latency reduction compared to spatial extensions, which we attribute to the strong inter-frame similarity inherent in video inputs.}
We find that a block size of 2×2×2 is sufficient to provide strong performance. 

\textbf{Scatter Accumulator.}
The number of accumulators in similarity scatter affects throughput and pipeline efficiency. Ideally, accumulation should finish before the next output tile arrives from the systolic array.
As shown in \Fig{fig:DSE}(d), using 64 accumulators achieves near-peak performance with only a 5\% latency overhead compared to a larger 160-accumulator design, with diminishing returns beyond that point. 
This configuration also simplifies buffer design.

{\textbf{Semantic Pruning Configuration.}
In our Semantic Pruning scheme, the value of “k” in top-k pruning is determined by multiplying the original number of image tokens by a predefined retention ratio. We search multiple layer-wise retention configurations and select the one offering the best sparsity–accuracy trade-off, which is adopted in our design. The final setup is summarized in \Tbl{tab:hardware-config}, where pruning is applied to five selected layers whose retention ratios differ from the preceding layer.
Future work may further enhance this strategy by dynamically adapting to input contexts, e.g., using a post-softmax attention threshold or top-p pruning~\cite{lin2025twilight}, though such adaptation can introduce runtime variations across inputs.
}

\subsection{Ablation Study}

{To assess the contribution of each component in \proj, we perform an ablation study on Llava-Video-7B and report speedup, as shown in \Fig{fig:ablation}. We incrementally enable the SEC and SIC, comparing results against a dense systolic baseline and CMC~\cite{song2024cmc}.
When only the SEC is enabled, \proj{} achieves a 3.15$\times$ speedup over the uncompressed systolic baseline and a \textbf{1.58$\times$} speedup over CMC. This demonstrates that semantic-aware pruning remove a large fraction of irrelevant visual tokens based on textual guidance, outperforming prior token-pruning strategies.

Enabling the vector-wise SIC further boosts speedup by an additional \textbf{1.44$\times$}. This highlights the ability of SIC to exploit residual redundancy among retained tokens at a finer vector granularity, beyond what semantic pruning alone can uncover.
SEC and SIC together yield a 4.53$\times$ speedup over the dense baseline and 2.26$\times$ over CMC, confirming the effectiveness and efficiency of the \proj{} design.}

\begin{figure}
    \centering
    \includegraphics[width=\linewidth]{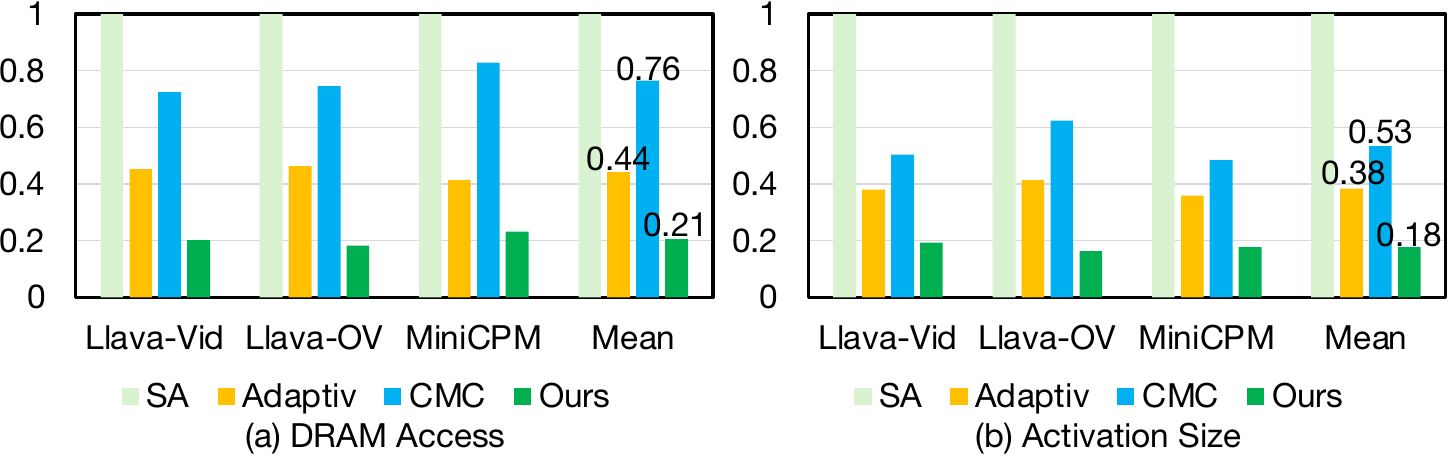}    
    \caption{Memory access analysis (a) overall DRAM access (b) activation size}
    \label{fig:mem_access}    
\end{figure}

\subsection{Memory Access}
\label{sec:mem_access}

We analyze the off-chip memory traffic and {average} input matrix size of \proj{} compared to baseline designs. As shown in \Fig{fig:mem_access}(a) and (b), \proj{} achieves the lowest DRAM access and input matrix size across all methods. Compared to the dense systolic array, we compress the input matrix by 5.6$\times$ and reduce memory traffic by 4.9$\times$.

This reduction stems from the joint effect of the SEC and SIC, which sparsifies the input at both token and vector levels. Additionally, the output of each FC layer is immediately compressed on-chip before being written to memory, so only compressed activations are transferred to DRAM, minimizing total memory access.

Compared to both CMC and AdapTiV, \proj{} achieves significantly higher input compression and lower DRAM access: \textbf{3.0$\times$} and \textbf{2.2$\times$} higher compression ratios, and \textbf{3.7$\times$} and \textbf{2.2$\times$} DRAM traffic reduction, respectively. 

CMC relies on codec-based similarity detection over wide temporal windows and full-token representations, requiring large uncompressed regions to be staged in DRAM before processing. This leads to redundant memory transfers, as data must be written and read again for similarity detection. Similarly, AdapTiV performs local token pruning but still processes on whole-token granularity.
By avoiding these limitations through lightweight, streaming-compatible similarity matching, \proj{} achieves superior memory efficiency with minimal overhead.

\subsection{Synergy with Quantization}
{\proj is fully compatible with standard quantization techniques. We integrate \proj{} with INT8 quantization using bitsandbytes~\cite{dettmers20218}, and the results in \Tbl{tab:int8_degrade} show the impact on accuracy and sparsity compared to FP16. INT8 causes an average accuracy drop of 0.5\% and a sparsity change of 0.13\% relative to FP16. Although this loss is slightly higher than the 0.02\% degradation in the dense model, it is reasonable since \proj and quantization jointly compress the model. Overall, the accuracy drop remains minor, and \proj effectively maintains its redundancy-removal capability under quantization, demonstrating strong synergy for efficient VLM inference.}

\begin{table}[t]
    \centering
    \caption{{Influence of INT8 quantization on accuracy and sparsity}}
    \resizebox{0.48\textwidth}{!}{
    \begin{tabular}{l|l|c|c|c|c|c|c}
    \toprule
        \multirow{2}{*}{Models} & \multirow{2}{*}{Datasets} 
        & \multicolumn{2}{c|}{Dense} 
        & \multicolumn{2}{c|}{Ours} 
        & \multicolumn{2}{c}{Ours} \\ 
        \cmidrule{3-8}
        ~ & ~ & Acc. & Degrade & Acc. & Degrade & Sparsity & Degrade \\ 
    \midrule
        \multirow{3}{*}{Llava-Vid} 
        & VMME & 64.22 & -0.07 & 62.33 & 0.41 & 82.48 & 0.34 \\ 
        & MLVU & 68.21 & -0.47 & 64.94 & 1.05 & 78.10 & 0.17 \\ 
        & MVB  & 59.75 &  0.58 & 57.95 & 0.25 & 78.04 & 0.40 \\ 
    \midrule
        \multirow{3}{*}{Llava-OV} 
        & VMME & 58.70 & -0.29 & 57.44 & 1.26 & 81.46 & 0.03 \\ 
        & MLVU & 63.38 & -0.06 & 62.41 & 0.11 & 78.35 & -0.01 \\ 
        & MVB  & 58.55 & -0.17 & 56.18 & 0.60 & 85.35 & 0.14 \\ 
    \midrule
        \multirow{3}{*}{MiniCPM} 
        & VMME & 58.63 &  0.18 & 57.96 & 0.34 & 82.84 & 0.03 \\ 
        & MLVU & 55.93 & -0.04 & 53.22 & 0.37 & 77.99 & 0.02 \\ 
        & MVB  & 55.13 &  0.50 & 54.03 & 0.27 & 75.97 & 0.02 \\ 
    \bottomrule
    \end{tabular}
    \label{tab:int8_degrade}
    }
\end{table}

\section{Discussion}
\label{sec:disc}

\subsection{Generalization Ability of \proj}
{We further examine the generalization ability of \proj. Although \textit{Focus} is originally designed for video-based VLMs, it can be directly extended to image-based VLMs by treating a single image as a one-frame video. While temporal similarity is no longer present in this setting, substantial semantic redundancy and spatial similarity remain. As shown in \Tbl{tab:image}, evaluations on image-based VLMs~\cite{li2024llava,bai2025qwen2} across multiple datasets~\cite{goyal2017making,liu2024mmbench,zhang2021mme} demonstrate notable speedups over both the systolic-array baseline and AdapTiV, with only minor accuracy degradation. These results indicate that \proj effectively removes redundancy beyond the video domain.}

Moreover, \proj can potentially be extended to Vision--Language--Action (VLA)~\cite{kim2024openvla,intelligence2504pi0} models for embodied AI applications. VLA models share similar input modalities with VLMs, including image or video inputs paired with text. Therefore, we believe the SEC and SIC in \proj could effectively eliminate redundant information in VLA inputs, making this a promising direction for future exploration.

\begin{table}[t]
    \centering
    \caption{Accuracy and Speedup on image VLMs}
    \resizebox{0.47\textwidth}{!}{
    \begin{tabular}{l|l|l|c|c|c}
    \toprule
        Models & Dataset & Metric & Dense & AdapTiV & Ours \\ \midrule
        \multirow{6}{*}{Llava-OneVision} 
        & \multirow{2}{*}{VQAv2} & Speedup & 1.00 & \textbf{5.19} & 4.44 \\ 
        & ~ & Accuracy & 84.32 & 82.48 & \textbf{83.01} \\ \cmidrule{2-6}
        & \multirow{2}{*}{MME} & Speedup & 1.00 & 1.65 & \textbf{4.26} \\ 
        & ~ & Score & 1067.27 & 1036.27 & \textbf{1044.79}  \\ \cmidrule{2-6}
        & \multirow{2}{*}{MMBench} & Speedup & 1.00 & 1.60 & \textbf{4.25} \\ 
        & ~ & Accuracy & 84.99 & \textbf{84.49} & 83.46 \\ \midrule
        \multirow{6}{*}{Qwen2.5-VL} 
        & \multirow{2}{*}{VQAv2} & Speedup & 1.00 & \textbf{1.96} & 1.91 \\ 
        & ~ & Accuracy & 84.48 & 79.77 & \textbf{81.73} \\ \cmidrule{2-6}
        & \multirow{2}{*}{MME} & Speedup & 1.00 & 1.89 & \textbf{1.97} \\ 
        & ~ & Score & 1337.66 & 1129.56 & \textbf{1238.88} \\ \cmidrule{2-6}
        & \multirow{2}{*}{MMBench} & Speedup & 1.00 & \textbf{1.93} & 1.78 \\ 
        & ~ & Accuracy & 85.69 & 83.79 & \textbf{84.46} \\
    \bottomrule
    \end{tabular}
    \label{tab:image}
    }    
\end{table}

\subsection{Worst- and Best-Case Analysis}
{Since sparsity varies with video content, we analyze two extreme scenarios to verify robustness.} {In the \textbf{worst case}, when no similarity exists across frames or patches, sparsity drops near zero. The design preserves the full tile length ($m=1024$), with buffers sized for maximum data without overflow.} {In the \textbf{best case}, abundant similarity yields highly compressed tiles and small $m$, slightly wasting buffer space and underutilizing the systolic array but remaining correct.} {We further aggregate the frequency of different tile lengths (i.e. number of vectors) and measure systolic-array utilization (\Fig{fig:worst_case}). These extremes are rare, one increases latency, the other lowers utilization, but the system maintains an average utilization of 92.2\%, confirming robust performance across diverse inputs.
}

\subsection{Related Works}

\textbf{Algorithm Optimizations.} 
As a major paradigm for multimodal reasoning, VLMs have attracted significant attention, resulting in a rapidly growing body of work on improving inference efficiency. 
A large class of recent methods focuses on exploiting redundancy in visual tokens to accelerate VLM inference~\cite{fu2024framefusion,fastv,huang2025prunevid,shen2025fastvid,liu2025keyframe,wang2025corematching,liu2025nvila}. 
FrameFusion~\cite{fu2024framefusion} compares and merges similar tokens across adjacent frames, while PruneVid~\cite{huang2025prunevid} performs token clustering and merges tokens within the same cluster. 
By removing or compressing redundant tokens through diverse algorithmic strategies, these approaches effectively reduce token counts and achieve notable speedups on GPUs.

Despite their effectiveness, these methods operate exclusively at the token level and are implemented as software-only optimizations. 
They are primarily designed for execution on general-purpose GPUs and do not consider how redundancy manifests at finer granularities or how it can be efficiently exploited from a bottom-up hardware perspective.

\begin{figure}[t]
    \centering
    \includegraphics[width=\linewidth]{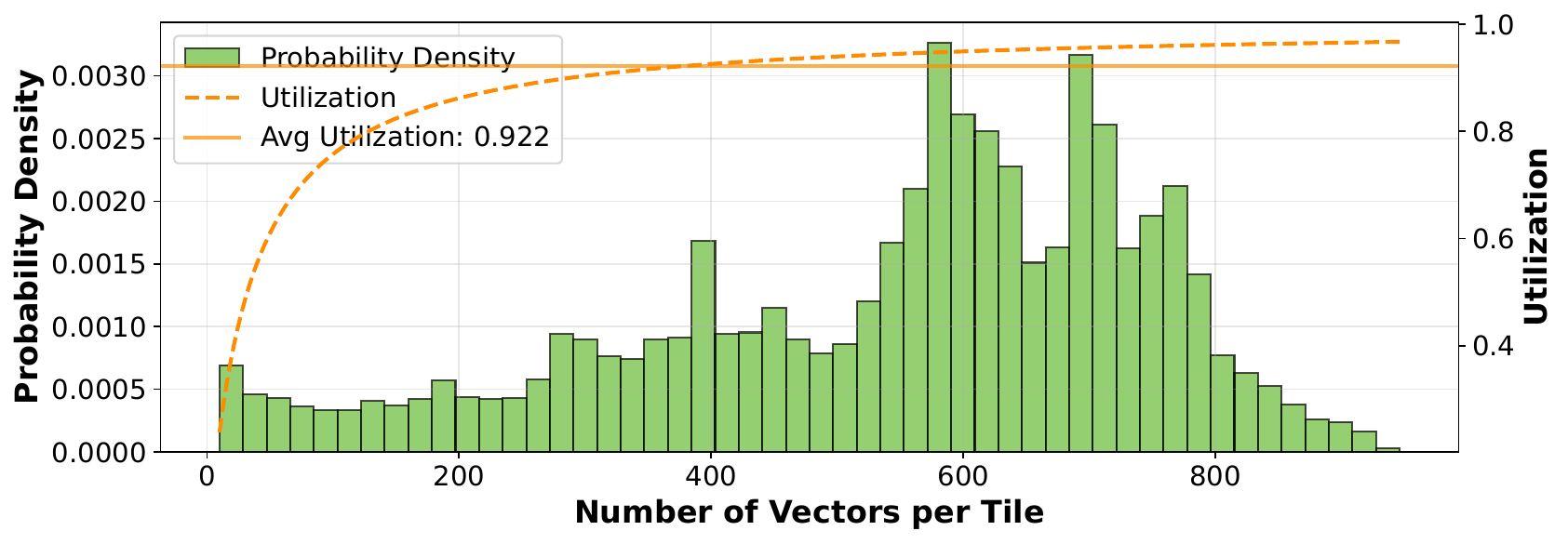}    
    \caption{{Histogram and compute utilization of concentrated tile length.} }
    \label{fig:worst_case}    
    \vspace{-2mm}
\end{figure}

\textbf{Architecture Design.} 
From the hardware architecture perspective, to the best of our knowledge, there is no dedicated accelerator specifically designed for VLM inference. 
Existing architectural efforts instead target efficiency optimizations for LLMs and ViTs, which share the transformer backbone with VLMs. 
These works predominantly rely on sparsity~\cite{wang2021spatten,dong2023heatvit,you2023vitcod,wang2025lad,guo2020accelerating,wei2025prosperity,wei2025phi} and quantization~\cite{guo2023olive,chen2025bitmod,guo2022ant,hu2025m,cheng2025ecco,guo2025transitive}.
In terms of sparsity, SpAtten~\cite{wang2021spatten} introduces token- and head-level pruning for transformers, while LAD~\cite{wang2025lad} optimizes key--value cache pruning during the decoding phase of LLMs. 
HeatViT~\cite{dong2023heatvit} leverages attention maps in ViTs to prune visual tokens. 
In addition, several works propose hardware-friendly quantization architectures: Olive~\cite{guo2023olive} introduces an outlier--victim pair format integrated into processing element (PE) arrays, and BitMoD~\cite{chen2025bitmod} enables fine-grained data-type adaptation through bit-serial processing.

While these techniques can be applied to VLMs, they are not explicitly designed for multimodal workloads and therefore may fail to fully capture the unique redundancy patterns introduced by cross-modal interactions. 
In contrast, \proj is the first architecture specifically tailored for VLM inference. 
By exploiting cross-modal semantic redundancy and detecting fine-grained vector-level similarity, \proj enables efficient streaming execution and achieves superior hardware efficiency beyond token-level or modality-agnostic optimizations.
 
\section{Conclusion}
We present \proj{}, a streaming concentration architecture that jointly optimizes algorithm and hardware for efficient VLM inference.  
Our Multilevel Concentration strategy removes redundancy at the semantic, block, and vector levels, while our hardware design performs in-place compression aligned with GEMM tiling and streaming execution.
\proj{} achieves up to \textbf{2.35$\times$} speedup and {\textbf{3.29$\times$}} energy efficiency improvement over state-of-the-art baselines, with only {\textbf{2.7\%}} area overhead in a systolic-array accelerator.
By tightly co-designing compression logic with accelerator architecture, \proj{} enables scalable, high-performance deployment of VLMs on both edge and cloud platforms, and paves the way for future hardware-aware multimodal systems.

\section*{Acknowledgment}

This work was supported in part by NSF-2112562, NSF-2328805, and ARO W911NF-23-2-0224. The authors sincerely thank the anonymous reviewers
for their constructive feedback and valuable suggestions that greatly
improved the quality of this work. The authors also express their gratitude to Jonathan Ku, Bowen Duan, Yiming Li, and Dr. Tingjun Chen for their technical support and insightful discussions.
%
%
%
%
%


\appendix
\section{Artifact Appendix}

\subsection{Abstract}

This artifact provides a complete implementation of \textit{Focus}, a streaming concentration architecture
for efficient vision-language model (VLM) inference. The artifact includes three main components:
(1) Algorithm implementation of Focus and baseline methods (CMC, Adaptiv, FrameFusion) for multiple VLMs, including LLaVA-Video, LLaVA-OneVision, MiniCPM-V, and Qwen2.5-VL; (2) Cycle-accurate hardware simulator with energy/power estimation and design space exploration capabilities; (3) RTL implementation in Verilog/SystemVerilog. The artifact enables the reproduction of all key results, including accuracy evaluations on the VideoMME, MLVU, MVBench, VQAv2, MME, and MMBench datasets, performance/energy simulations across various design configurations. Generated traces and simulation outputs can be used to reproduce all figures and tables in the evaluation section.

\subsection{Artifact check-list (meta-information)}

{\small
\begin{itemize}
  \item {\bf Algorithm: } Focus multilevel concentration (semantic, block, vector levels), CMC, Adaptiv, FrameFusion baselines
  \item {\bf Program: } Python 3.11+ with PyTorch 2.6.0+
  \item {\bf Compilation: } Python package installation via pip, third-party code compilation with g++
  \item {\bf Model: } LLaVA-Video-7B-Qwen2, MiniCPM-V-2.6, LLaVa-OneVision-qwen2-7b-ov, Qwen2.5-VL-7B-Instruct
  \item {\bf Dataset: } VideoMME, MLVU, MVBench (video); VQAv2, MME, MMBench (image)
  \item {\bf Run-time environment: } Ubuntu 22.04.2 LTS, CUDA 12.1, PyTorch 2.6.0, Conda environment, HuggingFace Hub access
  \item {\bf Hardware: } NVIDIA datacenter GPU (A100), multi-core CPU x86\_64 processor
  \item {\bf Execution: } Bash scripts for trace generation, Python scripts for simulation and evaluation, Jupyter notebooks for plotting
  \item {\bf Metrics: } Model accuracy, sparsity ratio, latency (cycles), energy (mJ), power (W), area (mm²)
  \item {\bf Output: } CSV files with accuracy/sparsity metrics, PyTorch trace files (.pth), simulation result CSVs, Jupyter notebooks for plotting
  \item {\bf Experiments: } Trace generation, accuracy evaluation, hardware simulation, design space exploration
  \item {\bf How much disk space required (approximately)?: } 128GB (models + datasets + traces + codes)
  \item {\bf How much time is needed to prepare workflow (approximately)?: } 1 hour (installation + model download)
  \item {\bf How much time is needed to complete experiments (approximately)?: } 6 hours without accuracy evaluation, 480 hours with accuracy evaluation.
  \item {\bf Publicly available?: } \href{https://github.com/dubcyfor3/Focus.git}{https://github.com/dubcyfor3/Focus.git}
  \item {\bf Code licenses (if publicly available)?: } MIT License
  \item {\bf Data licenses (if publicly available)?: } The datasets are publicly available through their original licensing terms.
  \item {\bf Workflow automation framework used?: } Bash scripts, Python entry points, Jupyter notebooks
  \item {\bf Archived (provide DOI)?: } \href{https://doi.org/10.5281/zenodo.17851347}{https://doi.org/10.5281/zenodo.17851347}
\end{itemize}
}

\subsection{Description}

\subsubsection{How to access}

The artifact is available as a Git repository at \href{https://github.com/dubcyfor3/Focus.git}{https://github.com/dubcyfor3/Focus.git}. Clone the repository and initialize submodules

\subsubsection{Hardware dependencies}

\begin{itemize}
  \item \textbf{GPU:} NVIDIA GPU with 80GB HBM (e.g. A100).
  \item \textbf{CPU:} x86\_64 processor
  \item \textbf{Storage:} 128GB+ available disk space
\end{itemize}

\subsubsection{Software dependencies}
The experiments rely on the following software components.
\begin{itemize}
  \item Ubuntu 22.04+ (tested on Ubuntu 22.04 LTS)
  \item Python 3.11+
  \item PyTorch 2.6.0 with CUDA support
  \item Transformers 4.48.2 (or 4.49.0 for Qwen2.5-VL)
  \item Accelerate 0.29.1+
  \item Flash-attention 2.7.4.post1
  \item g++ compiler
  \item HuggingFace CLI and account (for model/dataset access)
\end{itemize}

\subsubsection{Data sets} VideoMME, MLVU, MVBench, VQAv2, MME, MMBench

\subsubsection{Models}

The artifact supports multiple pre-trained VLMs accessible via HuggingFace:

\begin{itemize}
  \item \textbf{LLaVA-Video-7B-Qwen2} \\
  (\texttt{lmms-lab/LLaVA-Video-7B-Qwen2})
  \item \textbf{MiniCPM-V-2.6} \\
  (\texttt{openbmb/MiniCPM-V-2\_6})
  \item \textbf{LLaVA-OneVision} \\
  (\texttt{lmms-lab/llava-onevision-qwen2-7b-ov})
  \item \textbf{Qwen2.5-VL} \\
  (\texttt{Qwen/Qwen2.5-VL-7B-Instruct})
\end{itemize}

\subsection{Installation}

We have well-documented README files to detail the installation instructions for each experiment at \href{https://github.com/dubcyfor3/Focus.git}{https://github.com/dubcyfor3/Focus.git}

\subsection{Experiment workflow}

The README file also specifies the detailed experimental workflow for obtaining the results reported in the paper.

\subsection{Evaluation and expected results}

Comprehensive README files are provided to document the evaluation procedures for accelerator latency, energy, area, and model accuracy. Expected results can be found in the directories \texttt{simulator/example\_sim\_results} and \texttt{algorithm/example\_output}.

\subsection{Methodology}

Submission, reviewing, and badging methodology:

\begin{itemize}
  \item \url{https://www.acm.org/publications/policies/artifact-review-and-badging-current}
  \item \url{https://cTuning.org/ae}
\end{itemize}



\bibliographystyle{IEEEtranS}
\bibliography{refs}

\end{document}